\documentclass[%
reprint,
amsmath,amssymb,  
aps,
prl,
floatfix,
]{revtex4-2}

\usepackage{xcolor}
\usepackage[detect-all, abbreviations]{siunitx}
\usepackage{physics}
\usepackage{graphicx}
\usepackage{dcolumn}
\usepackage{bm}
\usepackage[colorlinks=true,linkcolor=blue,urlcolor=blue,citecolor=blue]{hyperref}
\usepackage[mathlines]{lineno}

\begin{document}
\newcommand{\Sr}{${}^{87}$Sr}
\newcommand{\Er}{$E_r$}


\preprint{}
\title{Evaluation of lattice light shift at low 10$^{-19}$ uncertainty for a shallow lattice Sr optical clock}
\author{Kyungtae Kim}
\affiliation{JILA, National Institute of Standards and Technology and University of Colorado \\ Department of Physics, University of Colorado, Boulder, Colorado 80309-0440, USA}
\author{Alexander Aeppli}
\affiliation{JILA, National Institute of Standards and Technology and University of Colorado \\ Department of Physics, University of Colorado, Boulder, Colorado 80309-0440, USA}
\author{Tobias Bothwell}
\affiliation{JILA, National Institute of Standards and Technology and University of Colorado \\ Department of Physics, University of Colorado, Boulder, Colorado 80309-0440, USA}
\author{Jun Ye}
\affiliation{JILA, National Institute of Standards and Technology and University of Colorado \\ Department of Physics, University of Colorado, Boulder, Colorado 80309-0440, USA}

\date{\today}

\begin{abstract}
A Wannier-Stark optical lattice clock has demonstrated unprecedented measurement precision for optical atomic clocks. We present a systematic evaluation of the lattice light shift, a necessary next step for establishing this system as an accurate atomic clock. With precise control of the atomic motional states in the lattice, we report accurate measurements of the multipolar and the hyperpolar contributions and the operational lattice light shift with a fractional frequency uncertainty of 3.5$\times$10${}^{-19}$.
\end{abstract}
\maketitle

\textbf{\emph{Introduction.}} Optical lattice clocks are advancing measurement precision to an unprecedented level~\cite{bothwellResolvingGravitationalRedshift2022,zhengDifferentialClockComparisons2022}. Achieving a similar level of measurement accuracy is both an expected natural development and a necessary condition for the future redefinition of time~\cite{
    ludlowOpticalAtomicClocks2015, 
    bothwellJILASrIOptical2019, 
    mcgrewAtomicClockPerformance2018, 
    letargatExperimentalRealizationOptical2013a, 
    boulderatomicclockopticalnetworkbaconcollaboration*FrequencyRatioMeasurements2021,
    hobsonStrontiumOpticalLattice2020, 
    nemitzFrequencyRatioYb2016, ushijimaCryogenicOpticalLattice2015, 
    schwarzLongTermMeasurement2020, 
    kimAbsoluteFrequencyMeasurement2021a}. 

The Sr optical lattice clock at JILA Sr1 employs a shallow one-dimensional (1D) optical lattice with enhanced atomic coherence and record low self-synchronous frequency instability~\cite{bothwellResolvingGravitationalRedshift2022}. This 1D lattice is established within an optical cavity oriented along the direction of gravity. Since neighboring sites are detuned by the gravitational potential energy difference, ultracold atoms confined in the lattice are described by Wannier-Stark (WS) wavefunctions~\cite{lemondeOpticalLatticeClock2005}. Important ingredients for such significant progress in clock precision include cooling a large yet dilute sample of fermionic $^{87}$Sr atoms to below 100~\si{nK}, well-characterized motional states, microscopic imaging spectroscopy, long atomic coherence time ($>$ 30~\si{s}), and the precise control of atomic interaction effects.  Further, at low lattice depths atom-atom interactions are modified, effectively eliminating density dependent frequency shifts~\cite{aeppliHamiltonianEngineeringSpinorbit2022}. Thus, a shallow, partially delocalized, WS optical lattice clock contains ideal characteristics for next generation timekeeping. 

\begin{figure}[h!]
    \includegraphics[width=\columnwidth]{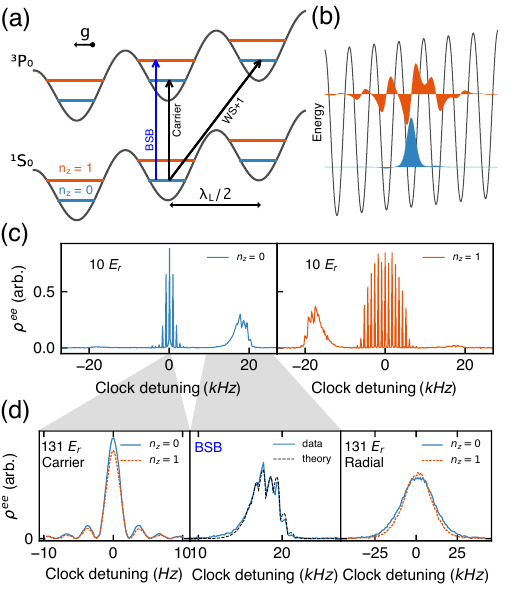}
    \caption{
        (a) Schematics of the 1D optical lattice system. Gravity $g$ lifts the energy degeneracy between adjacent lattice sites by $mg\lambda_{L}/2$. The clock carrier transition drives only $^1S_0$-$^3P_0$ without changing the axial quantum number, $n_z$, and is used to probe lattice light shift. For the axial state control, we use the blue sideband (BSB) to drive $\ket{n_z=0}$ to $\ket{n_z = 1}$. WS$\pm i$ denotes a transition to an $i$-site-shifted Wannier-Stark (WS) state.
        (b) Eigenstates of $n_z=0$ (blue) and $n_z=1$ (red) at $U_0=$ 15\Er{}. The offset of the wavefunction corresponds to eigenenergy of the lattice potential. (c-d) Characterization of the motional states. (c) The left(right) panel shows the axial sideband spectrum of $\ket{n_z=0}$($\ket{n_z=1}$). $\rho^{ee}$ is the excitation fraction. We adiabatically ramp $U_0$ to 10\Er{}, which supports only two axial states. Sidebands close to the carrier are WS$\pm i$ transitions. 
        (d) The left panel shows Rabi spectra of the two axial states. In the middle, we zoom-in to the motional sideband and plot with the theoretical lineshape~\cite{blattRabiSpectroscopyExcitation2009} taking into account WS$\pm i$ sidebands.  The right panel shows the Doppler broadening to extract $T_r$. For $\ket{n_z=0}$, $T_r$ is 500~\si{\nano\kelvin}. For $\ket{n_z=1}$, $T_r$ is 40\% lower, likely due to the limited transfer efficiency and reduced trapping potential.
        }
    \label{fig:1}
\end{figure}

While the spectroscopy lattice depth is far lower than previous clocks, the light shift associated with the lattice trapping light remains a key systematic. In this Letter, we report a detailed investigation of clock operation under engineered motional states within this titled lattice (Fig. \ref{fig:1}(a)). We provide a detailed study of lattice light shifts for lattice frequency, atomic motional states, and lattice depths near zero. In addition to reducing the total uncertainty of the lattice light shift down to $3.5\times 10^{-19}$ fractional frequency, we also report measurements of the light shift coefficients associated with the electric quadrupole (E2), magnetic dipole (M1) moments and the hyperpolarizability.

\textbf{\emph{Lattice light shift model.}} Before we set out to perform systematic measurements of the lattice light shift, we take important steps to reduce systematic effects. The cavity-based lattice establishes a stable and well-defined light mode and intensity calibrated directly to the trap depth. The density related frequency shift is reduced and precisely measured to remove the atomic interaction effects when the lattice depth is varied. The motional state and related transverse temperature are monitored with an independent probe, which is important for measuring the contribution from the E2/M1 term. 

The lattice light shift model is proposed for a lattice without considering the tunneling effect~\cite{ushijimaOperationalMagicIntensity2018, katoriStrategiesReducingLight2015, beloyModelingMotionalEnergy2020}. Since the gravitational tilt of the lattice is small compared to the bandgap, the model is valid with WS states~\cite{SeeSupplementalMaterial}. Here, we vary the lattice trap depth ranging from the WS regime to the more traditional isolated lattice configuration to explore the light shift effects. 

The lattice light shift of the clock transition, $\Delta \nu_{LS}$, can be expressed as a function of three control parameters; the lattice depth, $u$, lattice frequency, $\nu_L$, and the axial state quantum number, $n_z$. Following the convention established in~\cite{ushijimaOperationalMagicIntensity2018, katoriStrategiesReducingLight2015}, the lattice light shift can be written as
\begin{equation}
    \begin{aligned}    
        h \Delta \nu_{LS} &( u, \delta_L, n_z ) \\ 
        \approx& \left( \frac{\partial \tilde{\alpha}^{E1}}{\partial \nu} \delta_L - \tilde{\alpha}^{qm} \right)\left(n_z + \frac{1}{2}\right) u^{1/2} \\
        &- \left[ \frac{\partial \tilde{\alpha}^{E1}}{\partial \nu} \delta_L + \frac{3}{2}\tilde{\beta} \left( n_z^2 + n_z + \frac{1}{2}\right)\right] u^1 \\
        &+ 2\tilde{\beta} \left(n_z + \frac{1}{2}\right)u^{3/2} - \tilde{\beta} u^2.
        \label{eq:light_shift}
    \end{aligned}
\end{equation}
The transverse motional effect is accounted for with the use of an effective depth, $u^j = (1+jk_BT_r/u_0E_r)^{-1}u_0^{j}$, where $k_B$ is the Boltzmann constant, $T_r$ the radial temperature measured by transverse Doppler spectroscopy (Fig.~\ref{fig:1}(d)), $u_0 = U_0/E_r$ is the peak lattice depth, and the superscript $j$ represents the thermally averaged $j$th power of the lattice depth~\cite{beloyModelingMotionalEnergy2020}. $E_r = h^2/2m\lambda_{L}^2$, where $m$ is the mass of $^{87}$Sr and $\lambda_{L}$ = $c/\nu_L$ the lattice wavelength, $h$ is the Planck constant, with $c$ the speed of light. $\delta_L = \nu_L - \nu^{E1}$ is the detuning from E1 magic frequency ($\nu^{E1}$), including both the differential scalar shift and an experiment-specific differential tensor shift.

There are four dependent coefficients to be characterized, including $\nu^{E1}$. ${\partial \tilde{\alpha}^{E1}}/{\partial \nu}$ is the frequency derivative of the differential electric dipole (E1) polarizability between $\ket{{}^1S_0, m_F =  \pm\frac{5}{2}}$ and $\ket{{}^3P_0, m_F = \pm\frac{3}{2}}$ near $\nu^{E1}$. This term includes both the scalar and tensor contributions, while the vector contribution is cancelled by clock interrogation of opposite sign $m_F$ states. $\tilde{\alpha}^{qm}/h$ is the differential multipolar polarizability in units of \si{\hertz}, and $\tilde{\beta}/h$ is the differential hyperpolarizability in units of \si{\hertz}. The variation of $n_z$ is critical for enhancing the sensitivity to $\tilde{\alpha}^{qm}/h$, because different wavefunction extensions of $\ket{n_z=1}$ and $\ket{n_z=0}$ (Fig. \ref{fig:1}(b)) vary the weighting factor between E1 (maximum at the lattice anti-node) and E2-M1 (maximum at the node).

We systematically explore Eq.~\eqref{eq:light_shift} by measuring the frequency difference between two control parameter sets. We vary the lattice depth from 3\Er{} to 300\Er{}, $\delta_L$ over $\pm200$~\si{MHz}, and $n_z$ = 0 and 1. When we modulate $u$ and $\delta_L$, the reference is chosen to be at the magic lattice depth to suppress the systematic error from collisional shifts~\cite{aeppliHamiltonianEngineeringSpinorbit2022}. We do not investigate the separation of tensor and scalar polarizability because they are integrated into ${\partial \tilde{\alpha}^{E1}}/{\partial \nu}$. 

\textbf{\emph{Clock in tilted, shallow lattice.}} Our 1D \Sr~optical lattice clock is detailed in previous publications~\cite{aeppliHamiltonianEngineeringSpinorbit2022,bothwellResolvingGravitationalRedshift2022}. We prepare stretched states ($m_F = \pm 9/2$) spin-polarized ensembles in a single motional ground state axially and $T_r\sim700$~\si{\nano\kelvin} radially at $U_0 = $ 300\Er{}. The lattice intensity is adiabatically ramped to a range of depths and $T_r$ is confirmed to vary from 700~\si{nK} to 60~\si{nK}. The atom number is about $10^5$ for ($u$, $\nu_L$) dependence measurement and $2\times 10^4$ for $n_z$ dependence measurement. Fig.~\ref{fig:1}(c) presents a spectroscopic characterization of the motional state distribution of the atoms. After the preparation, we lower the lattice depth to the desired level and transfer the spin states with a series of clock $\pi$-pulses together with cleaning pulses. In all cases, we use the magnetically insensitive $\ket{{}^1S_0, m_F = \pm\frac{5}{2}}\rightarrow\ket{{}^3P_0, m_F = \pm \frac{3}{2}}$ transition. For axial state control, a clock pulse resonant to the blue sideband (shown in Fig.~\ref{fig:1}(a)) drives $\ket{n_z=0} \rightarrow \ket{n_z=1}$ at $U_0= $ 22\Er{}. Transfer efficiency is about 15-20\% and results in a $T_r$ reduction of 40~\% (see Fig.~\ref{fig:1}(d)). Nevertheless, this temperature difference is negligible due to the low temperature and the proximity to the E1 magic frequency. 

We use a cryogenic silicon cavity-stabilized laser to drive the clock transition~\cite{oelkerDemonstration10172019,mateiMmLasersSub102017}.  Two interleaved atomic servos at two different conditions ($u$, $\delta_L$, $n_z$) track the clock resonance and continuously average the differential frequency shift~\cite{nicholsonComparisonTwoIndependent2012}. With a cavity stability of $4\times10^{-17}$, the Dick effect limited self-comparison stability is about $2\times 10^{-16}$ at 1~\si{\second} with 380~\si{\milli\second} Rabi pulse and about $1~\si{s}$ duty cycle. We determine the frequency shift by averaging collected frequency differences and assign 1$\sigma$ statistical uncertainty from a fit to the overlapping Allan deviation taken at 1/3 of the total measurement time $\tau$. In all cases, density shift corrected Allan deviations of the frequency difference follow the expected white frequency noise trend of $1/\sqrt{\tau}$. Typical uncertainties are less than $3\times10^{-18}$ for ($u$, $\delta_L$) modulation and about $5\times10^{-18}$ for $n_z$ modulation.

We apply a 70~\si{\micro\tesla} bias magnetic field during the clock interrogation. The direction is parallel to the polarization of the lattice laser to minimize sensitivity to the polarization fluctuation~\cite{westergaardLatticeInducedFrequencyShifts2011,shiPolarizabilitiesSr872015}. The vector shift and the field fluctuations are corrected as we interrogate two opposite spins of the atoms. After the clock interrogation, we ramp the lattice up to 300\Er{} and measure the excitation fraction with a standard shelving technique. The camera readout provides a high-resolution spatial distribution of the density and excitation fraction. With this information, we correct the density shift shot-by-shot~\cite{aeppliHamiltonianEngineeringSpinorbit2022}, providing a robust rejection of systematics related to the atomic density fluctuation.

To establish the lattice, we seed the in-vacuum cavity with an injection-seeded diode laser ($<$ 500~\si{\milli\watt}) to reach a lattice depth up to 300\Er{} with the waist, $w_0$ of 260~\si{\micro\meter}. A volume Bragg grating with 50~\si{\giga\hertz} bandwidth and the optical cavity finesse of 1000 greatly suppress the broad spectral background of the diode laser~\cite{fasanoCharacterizationSuppressionBackground2021}. With the lattice laser frequency locked to a cavity resonance, the cavity itself is stabilized to an absolute frequency-stabilized optical frequency comb. For lattice frequency modulation, we vary a comb-lock offset frequency so the cavity continuously follows the laser during the sample preparation before the last cooling stage. This scheme allows us to change the lattice frequency by $\pm$200~\si{\mega\hertz} within 200~\si{\milli\second}, suitable for interleaved self-comparison. 

\begin{figure}[h]
    \includegraphics[width=\columnwidth]{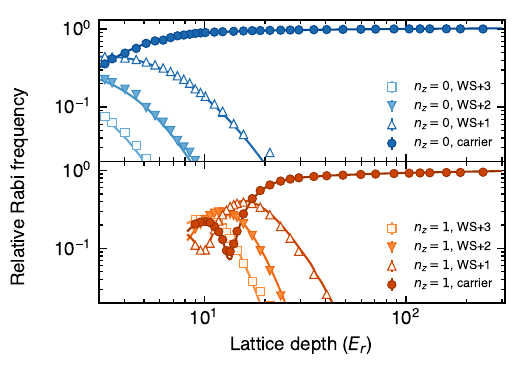}
    \centering
    \caption{
        Rabi frequency for the carrier and WS$+i$ transitions. Measured Rabi frequencies are normalized to the maximum value. The upper (lower) panel shows WS$+i$ transitions of $\ket{n_z=0}$ ($\ket{n_z=1}$). The error bars indicate 1 standard deviation from the fitting. Solid lines are theory calculations. The lattice depth, $U_0$ is fit to this data for the calibration. As we reduce $U_0$, atoms are delocalized and their coupling to the clock laser drive is reduced significantly. 
        }
    \label{fig:2}
\end{figure}

\textbf{\emph{Rabi excitation of Wannier-Stark states.}} 
Understanding atom-laser interaction at the shallow lattice depth is essential for the light shift evaluation. The tunneling rate between the lattice sites is exponentially sensitive to $u$. Hence, for small $u$, the extent of the delocalized atomic wavefunction can be larger than the clock laser wavelength (Fig.~\ref{fig:1}(b)), resulting in a breakdown of the Lamb-Dicke regime and a dramatic reduction of the clock drive Rabi frequency. We note that the long atom-light coherence is critically important here as it ensures a resolved sideband regime for spectroscopy. A time-domain Rabi oscillation signal is fit with a decayed sinusoidal curve to extract the Rabi frequency~\cite{SeeSupplementalMaterial}. Relative Rabi frequencies of the carrier and three site-changing transitions (WS$+i$) are shown in Fig.~\ref{fig:2} for the $n_z = 0$ and $n_z = 1$ motional states. Solid lines are numerically calculated Rabi frequencies fitted to the data with two fitting parameters: the cavity-transmitted light intensity conversion factor to lattice depth and an overall normalization factor for Rabi frequency. We calibrate the peak lattice depth, $U_0$, based on this fit, and the uncertainty is $0.1$($0.6$)\Er{} at $U_0=$ 0(300)\Er{}.

The increased sensitivity of the Rabi frequency on the lattice depth is reflected in its radial variation. Consequently, coupling to the second order radial sidebands becomes more pronounced, especially for $\ket{n_z = 1}$. See Supplementary Material for the details~\cite{SeeSupplementalMaterial}.

\textbf{\emph{Deviation from the stationary state.}} 
At very low lattice depths ($\leq 6E_r$), stationary Wannier-Stark states may no longer be supported, as evidenced by the increasingly large effective tunneling rate in Fig. S8 of Supplementary Material~\cite{SeeSupplementalMaterial}. We observe a deviation of the measured data from the lattice light shift model, with a rapid deterioration of the model fit if we include increasingly low lattice depth data~\cite{SeeSupplementalMaterial}. 

This deviation depends on the lattice depth $U$ and is insensitive to $\delta_L$. We vary other experimental conditions such as $\pi$-pulse duration, changing the initial state to $^3P_0$, and the bias field strength, with no effect on the deviation at the lowest trap depths. Therefore, we conclude that it is not from the lattice light shift. We compare two $\pi$-pulse durations that differ by factor 3 and observe no difference, suggesting the line pulling effect from the Bloch oscillation or the superposition of different WS states do not contribute to the deviation.

To add experimental evidence to the underlying mechanism of this deviation, we compare the clock frequency between the upward and downward propagation direction under otherwise the same condition~\cite{SeeSupplementalMaterial}. We find that the frequency deviation from the model using the opposite clock laser propagation has the same magnitude and opposite sign. Based on these observations, we exclude data where $U_0< 8 $\Er{} for $(u, \delta_L)$ modulation and $U_0< 30$\Er{} for $n_z$ modulation from the fit. 

\begin{figure}[h]
    \includegraphics[width=\columnwidth]{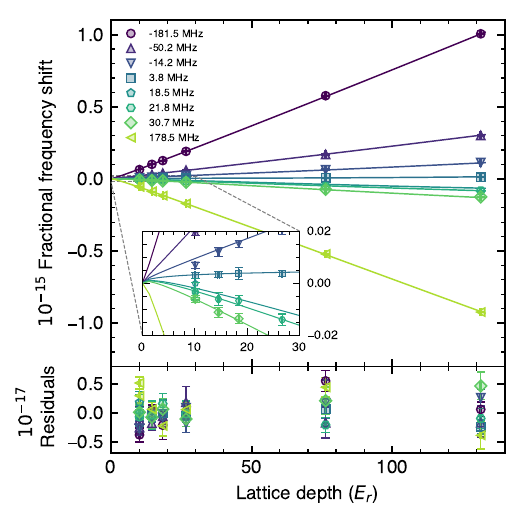}
    \centering
    \caption{ Lattice depth ($U_0$) and detuning ($\delta_L$) dependent light shift. Different markers represent different $\delta_L$ from the fit. For visualization, we offset the reference condition of each data point using the global fitting result. The inset (zoom-in) emphasizes a nonlinearity near the zero depth. The solid lines fit the model \eqref{eq:light_shift}. The fitting result is summarized in Table~\ref{tab:1}. We excluded data for $< 8E_r$~(see text for the details). The error bars show the 1$\sigma$ of statistical uncertainties. The shift uncertainty is calculated at 1/3 total measurement time using a $1/\sqrt{\tau}$ fit to the overlapping Allan deviation, and the depth uncertainties are from the lattice depth calibration.
    }
    \label{fig:3}
\end{figure}

\textbf{\emph{Lattice light shift evaluation.}} 
We explore Eq.~\eqref{eq:light_shift} in a self-contained manner with all three control parameters, $(u, \delta_L, n_z)$ and extract all the coefficients simultaneously. The results are summarized in Table~\ref{tab:1}. The residual variance (reduced $\chi^2$) of the fit is 1.3. The overall uncertainty includes inflation by the square root of the reduced $\chi^2$ and other systematic uncertainties from $T_r$.  For an operational condition, [$u = 10(0.2), \delta_L=0(0.1)~\si{MHz}, n_z=0(0.03)$], the uncertainty of the lattice light shift is $3.5\times10^{-19}$.





\begin{table}
    \centering
    \def\arraystretch{1.3}%
    \begin{tabular}{ c c c }
        \hline\hline
        Quantity &  Value\\
        \hline
        $\partial_{\nu} \tilde{\alpha}^{E1}/ h$             &  $1.859(5)\times 10^{-11}$         \\
        $\nu^{E1}$ (\si{\mega\hertz})                         & $368,554,825.9(4)$                 \\
        $\tilde{\alpha}^{qm}/h$ (\si{\milli\hertz})         & $-1.24(5)$                                   \\
        $\tilde{\beta}/h$     (\si{\micro\hertz})             & $-0.51(4)$                                  \\
        \hline\hline
    \end{tabular}
    \caption{
        Summary of the light shift characterization. We perform a single fit to the data including both Fig.~\ref{fig:3}, \ref{fig:4} to extract the coefficients.  For an operational condition, [$u = 10(0.2), \delta_L=0(0.1)~\si{MHz}, n_z=0(0.03)$], the uncertainty of the lattice light shift is $3.5\times10^{-19}$.
        }
    \label{tab:1}
\end{table}

Figure~\ref{fig:3} displays the light shift measurement investigating the dependence on $u$ and $\delta_L$. For this plot, only the carrier transition for $\ket{n_z=0}$ is employed. By choosing the reference condition near the magic lattice depth (10\Er{}) where the atomic density effect is suppressed~\cite{aeppliHamiltonianEngineeringSpinorbit2022, bothwellResolvingGravitationalRedshift2022}, we establish a reference frequency with minimal potential systematic effects. 

\begin{figure}[h]
    \includegraphics[width=\columnwidth]{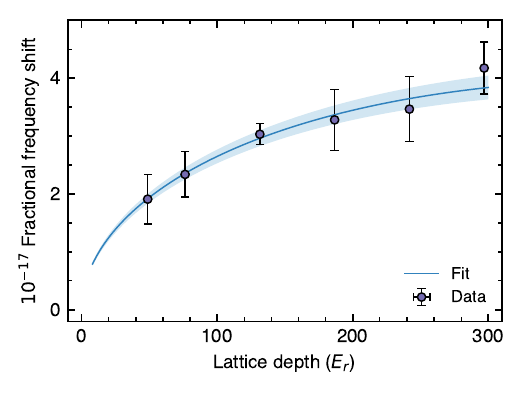}
    \centering
    \caption{Axial state dependent light shift, $\Delta \nu_{LS}(u, \delta_L, n_z=1) - \Delta \nu_{LS}(u, \delta_L, n_z=0) $, where $\delta_L <1~\si{MHz}$. We plot $n_z$ modulated part of the data set with good sensitivity to the multipolar polarizability, $\tilde{\alpha}^{qm}$. The solid line is fitting to the model \eqref{eq:light_shift} and the shades show 1$\sigma$ deviation of  $\tilde{\alpha}^{qm}$. Data points with $U_0<30E_r$ are excluded from the fit (see text for the details). The shift uncertainty is calculated at 1/3 total measurement time using a $1/\sqrt{\tau}$ fit to the overlapping Allan deviation.
    }
    \label{fig:4}
\end{figure}
Figure~\ref{fig:4} shows $n_z$ dependent light shift. For this part of data, we keep the other control parameters, ($u, \delta_L$) the same. We determine $\tilde{\alpha}^{qm} = -1.24(5)~\si{mHz}$. This value is close to recent measurements in deep lattices~\cite{dorscherExperimentalDeterminationE2M12022, ushijimaOperationalMagicIntensity2018}.A difference from \cite{ushijimaOperationalMagicIntensity2018, dorscherExperimentalDeterminationE2M12022} is that we use magnetic field insensitive spin states and microscopically determined $n_z$-dependent density shift coefficients ($\sim$ 20\% difference between $n_z =0, 1$) to suppress systematics~\cite{bothwellResolvingGravitationalRedshift2022,aeppliHamiltonianEngineeringSpinorbit2022}. While theory predictions still have disagreements ~\cite{porsevMultipolarPolarizabilitiesHyperpolarizabilities2018,wuDynamicMultipolarPolarizabilities2019,ovsiannikovHigherorderEffectsPrecision2016, ovsiannikovMultipoleNonlinearAnharmonic2013, katoriStrategiesReducingLight2015}, we became aware of a new theoretical calculation of $\tilde{\alpha}^{qm}$~\cite{wuContributionsNegativeenergyStates2023} that is in agreement with experimental observations.  The measured $\tilde{\beta}$ has a negligible contribution from the coupling of vector and tensor polarizability~\cite{shiPolarizabilitiesSr872015}, and it agrees with the previous measurements~\cite{porsevMultipolarPolarizabilitiesHyperpolarizabilities2018,ushijimaOperationalMagicIntensity2018,nicholsonSystematicEvaluationAtomic2015,letargatExperimentalRealizationOptical2013a,westergaardLatticeInducedFrequencyShifts2011}.

\textbf{\emph{Conclusion.}} With precise control of the motional states in a Wannier-Stark optical lattice, we show that an important systematic effect, the lattice light shift, is measured and controlled at the 3.5$\times$10$^{-19}$ uncertainty. This result is unique in that the light shift model is tested for very shallow lattices. This is important for achieving high accuracy optical lattice clocks for the future definition of SI Second. 

\textbf{\emph{Acknowledgements.}} We thank D. Kedar and C. Kennedy for technical discussions and assistance. We thank A. Chu and W. McGrew for careful reading of the manuscript and useful comments. Funding support is provided by NSF QLCI OMA-2016244, DOE Center of Quantum System Accelerator, DARPA, NIST, and NSF Phys-1734006. K. K. was supported by the education and training program of the Quantum Information Research Support Center, funded through the National Research Foundation of Korea(NRF) by the Ministry of science and ICT(MSIT) of the Korean government(No.2021M3H3A103657313).

\bibliography{main}

\begin{thebibliography}{32}%
\makeatletter
\providecommand \@ifxundefined [1]{%
 \@ifx{#1\undefined}
}%
\providecommand \@ifnum [1]{%
 \ifnum #1\expandafter \@firstoftwo
 \else \expandafter \@secondoftwo
 \fi
}%
\providecommand \@ifx [1]{%
 \ifx #1\expandafter \@firstoftwo
 \else \expandafter \@secondoftwo
 \fi
}%
\providecommand \natexlab [1]{#1}%
\providecommand \enquote  [1]{``#1''}%
\providecommand \bibnamefont  [1]{#1}%
\providecommand \bibfnamefont [1]{#1}%
\providecommand \citenamefont [1]{#1}%
\providecommand \href@noop [0]{\@secondoftwo}%
\providecommand \href [0]{\begingroup \@sanitize@url \@href}%
\providecommand \@href[1]{\@@startlink{#1}\@@href}%
\providecommand \@@href[1]{\endgroup#1\@@endlink}%
\providecommand \@sanitize@url [0]{\catcode `\\12\catcode `\$12\catcode
  `\&12\catcode `\#12\catcode `\^12\catcode `\_12\catcode `\%12\relax}%
\providecommand \@@startlink[1]{}%
\providecommand \@@endlink[0]{}%
\providecommand \url  [0]{\begingroup\@sanitize@url \@url }%
\providecommand \@url [1]{\endgroup\@href {#1}{\urlprefix }}%
\providecommand \urlprefix  [0]{URL }%
\providecommand \Eprint [0]{\href }%
\providecommand \doibase [0]{https://doi.org/}%
\providecommand \selectlanguage [0]{\@gobble}%
\providecommand \bibinfo  [0]{\@secondoftwo}%
\providecommand \bibfield  [0]{\@secondoftwo}%
\providecommand \translation [1]{[#1]}%
\providecommand \BibitemOpen [0]{}%
\providecommand \bibitemStop [0]{}%
\providecommand \bibitemNoStop [0]{.\EOS\space}%
\providecommand \EOS [0]{\spacefactor3000\relax}%
\providecommand \BibitemShut  [1]{\csname bibitem#1\endcsname}%
\let\auto@bib@innerbib\@empty
\bibitem [{\citenamefont {Bothwell}\ \emph {et~al.}(2022)\citenamefont
  {Bothwell}, \citenamefont {Kennedy}, \citenamefont {Aeppli}, \citenamefont
  {Kedar}, \citenamefont {Robinson}, \citenamefont {Oelker}, \citenamefont
  {Staron},\ and\ \citenamefont
  {Ye}}]{bothwellResolvingGravitationalRedshift2022}%
  \BibitemOpen
  \bibfield  {author} {\bibinfo {author} {\bibfnamefont {T.}~\bibnamefont
  {Bothwell}}, \bibinfo {author} {\bibfnamefont {C.~J.}\ \bibnamefont
  {Kennedy}}, \bibinfo {author} {\bibfnamefont {A.}~\bibnamefont {Aeppli}},
  \bibinfo {author} {\bibfnamefont {D.}~\bibnamefont {Kedar}}, \bibinfo
  {author} {\bibfnamefont {J.~M.}\ \bibnamefont {Robinson}}, \bibinfo {author}
  {\bibfnamefont {E.}~\bibnamefont {Oelker}}, \bibinfo {author} {\bibfnamefont
  {A.}~\bibnamefont {Staron}},\ and\ \bibinfo {author} {\bibfnamefont
  {J.}~\bibnamefont {Ye}},\ }\bibfield  {title} {\bibinfo {title} {Resolving
  the gravitational redshift across a millimetre-scale atomic sample},\ }\href
  {https://doi.org/10.1038/s41586-021-04349-7} {\bibfield  {journal} {\bibinfo
  {journal} {Nature}\ }\textbf {\bibinfo {volume} {602}},\ \bibinfo {pages}
  {420} (\bibinfo {year} {2022})}\BibitemShut {NoStop}%
\bibitem [{\citenamefont {Zheng}\ \emph {et~al.}(2022)\citenamefont {Zheng},
  \citenamefont {Dolde}, \citenamefont {Lochab}, \citenamefont {Merriman},
  \citenamefont {Li},\ and\ \citenamefont
  {Kolkowitz}}]{zhengDifferentialClockComparisons2022}%
  \BibitemOpen
  \bibfield  {author} {\bibinfo {author} {\bibfnamefont {X.}~\bibnamefont
  {Zheng}}, \bibinfo {author} {\bibfnamefont {J.}~\bibnamefont {Dolde}},
  \bibinfo {author} {\bibfnamefont {V.}~\bibnamefont {Lochab}}, \bibinfo
  {author} {\bibfnamefont {B.~N.}\ \bibnamefont {Merriman}}, \bibinfo {author}
  {\bibfnamefont {H.}~\bibnamefont {Li}},\ and\ \bibinfo {author}
  {\bibfnamefont {S.}~\bibnamefont {Kolkowitz}},\ }\bibfield  {title} {\bibinfo
  {title} {Differential clock comparisons with a multiplexed optical lattice
  clock},\ }\href {https://doi.org/10.1038/s41586-021-04344-y} {\bibfield
  {journal} {\bibinfo  {journal} {Nature}\ }\textbf {\bibinfo {volume} {602}},\
  \bibinfo {pages} {425} (\bibinfo {year} {2022})}\BibitemShut {NoStop}%
\bibitem [{\citenamefont {Ludlow}\ \emph {et~al.}(2015)\citenamefont {Ludlow},
  \citenamefont {Boyd}, \citenamefont {Ye}, \citenamefont {Peik},\ and\
  \citenamefont {Schmidt}}]{ludlowOpticalAtomicClocks2015}%
  \BibitemOpen
  \bibfield  {author} {\bibinfo {author} {\bibfnamefont {A.~D.}\ \bibnamefont
  {Ludlow}}, \bibinfo {author} {\bibfnamefont {M.~M.}\ \bibnamefont {Boyd}},
  \bibinfo {author} {\bibfnamefont {J.}~\bibnamefont {Ye}}, \bibinfo {author}
  {\bibfnamefont {E.}~\bibnamefont {Peik}},\ and\ \bibinfo {author}
  {\bibfnamefont {P.~O.}\ \bibnamefont {Schmidt}},\ }\bibfield  {title}
  {\bibinfo {title} {Optical atomic clocks},\ }\href
  {https://doi.org/10.1103/RevModPhys.87.637} {\bibfield  {journal} {\bibinfo
  {journal} {Reviews of Modern Physics}\ }\textbf {\bibinfo {volume} {87}},\
  \bibinfo {pages} {637} (\bibinfo {year} {2015})}\BibitemShut {NoStop}%
\bibitem [{\citenamefont {Bothwell}\ \emph {et~al.}(2019)\citenamefont
  {Bothwell}, \citenamefont {Kedar}, \citenamefont {Oelker}, \citenamefont
  {Robinson}, \citenamefont {Bromley}, \citenamefont {Tew}, \citenamefont
  {Ye},\ and\ \citenamefont {Kennedy}}]{bothwellJILASrIOptical2019}%
  \BibitemOpen
  \bibfield  {author} {\bibinfo {author} {\bibfnamefont {T.}~\bibnamefont
  {Bothwell}}, \bibinfo {author} {\bibfnamefont {D.}~\bibnamefont {Kedar}},
  \bibinfo {author} {\bibfnamefont {E.}~\bibnamefont {Oelker}}, \bibinfo
  {author} {\bibfnamefont {J.~M.}\ \bibnamefont {Robinson}}, \bibinfo {author}
  {\bibfnamefont {S.~L.}\ \bibnamefont {Bromley}}, \bibinfo {author}
  {\bibfnamefont {W.~L.}\ \bibnamefont {Tew}}, \bibinfo {author} {\bibfnamefont
  {J.}~\bibnamefont {Ye}},\ and\ \bibinfo {author} {\bibfnamefont {C.~J.}\
  \bibnamefont {Kennedy}},\ }\bibfield  {title} {\bibinfo {title} {{{JILA SrI}}
  optical lattice clock with uncertainty of 2.0 \texttimes{}
  10{\textsuperscript{-18}}},\ }\href
  {https://doi.org/10.1088/1681-7575/ab4089} {\bibfield  {journal} {\bibinfo
  {journal} {Metrologia}\ }\textbf {\bibinfo {volume} {56}},\ \bibinfo {pages}
  {065004} (\bibinfo {year} {2019})},\ \Eprint
  {https://arxiv.org/abs/1906.06004} {arXiv:1906.06004} \BibitemShut {NoStop}%
\bibitem [{\citenamefont {McGrew}\ \emph {et~al.}(2018)\citenamefont {McGrew},
  \citenamefont {Zhang}, \citenamefont {Fasano}, \citenamefont {Sch{\"a}ffer},
  \citenamefont {Beloy}, \citenamefont {Nicolodi}, \citenamefont {Brown},
  \citenamefont {Hinkley}, \citenamefont {Milani}, \citenamefont {Schioppo},
  \citenamefont {Yoon},\ and\ \citenamefont
  {Ludlow}}]{mcgrewAtomicClockPerformance2018}%
  \BibitemOpen
  \bibfield  {author} {\bibinfo {author} {\bibfnamefont {W.~F.}\ \bibnamefont
  {McGrew}}, \bibinfo {author} {\bibfnamefont {X.}~\bibnamefont {Zhang}},
  \bibinfo {author} {\bibfnamefont {R.~J.}\ \bibnamefont {Fasano}}, \bibinfo
  {author} {\bibfnamefont {S.~A.}\ \bibnamefont {Sch{\"a}ffer}}, \bibinfo
  {author} {\bibfnamefont {K.}~\bibnamefont {Beloy}}, \bibinfo {author}
  {\bibfnamefont {D.}~\bibnamefont {Nicolodi}}, \bibinfo {author}
  {\bibfnamefont {R.~C.}\ \bibnamefont {Brown}}, \bibinfo {author}
  {\bibfnamefont {N.}~\bibnamefont {Hinkley}}, \bibinfo {author} {\bibfnamefont
  {G.}~\bibnamefont {Milani}}, \bibinfo {author} {\bibfnamefont
  {M.}~\bibnamefont {Schioppo}}, \bibinfo {author} {\bibfnamefont {T.~H.}\
  \bibnamefont {Yoon}},\ and\ \bibinfo {author} {\bibfnamefont {A.~D.}\
  \bibnamefont {Ludlow}},\ }\bibfield  {title} {\bibinfo {title} {Atomic clock
  performance enabling geodesy below the centimetre level},\ }\href
  {https://doi.org/10.1038/s41586-018-0738-2} {\bibfield  {journal} {\bibinfo
  {journal} {Nature}\ }\textbf {\bibinfo {volume} {564}},\ \bibinfo {pages}
  {87} (\bibinfo {year} {2018})}\BibitemShut {NoStop}%
\bibitem [{\citenamefont {Le~Targat}\ \emph {et~al.}(2013)\citenamefont
  {Le~Targat}, \citenamefont {Lorini}, \citenamefont {Le~Coq}, \citenamefont
  {Zawada}, \citenamefont {Gu{\'e}na}, \citenamefont {Abgrall}, \citenamefont
  {Gurov}, \citenamefont {Rosenbusch}, \citenamefont {Rovera}, \citenamefont
  {Nag{\'o}rny}, \citenamefont {Gartman}, \citenamefont {Westergaard},
  \citenamefont {Tobar}, \citenamefont {Lours}, \citenamefont {Santarelli},
  \citenamefont {Clairon}, \citenamefont {Bize}, \citenamefont {Laurent},
  \citenamefont {Lemonde},\ and\ \citenamefont
  {Lodewyck}}]{letargatExperimentalRealizationOptical2013a}%
  \BibitemOpen
  \bibfield  {author} {\bibinfo {author} {\bibfnamefont {R.}~\bibnamefont
  {Le~Targat}}, \bibinfo {author} {\bibfnamefont {L.}~\bibnamefont {Lorini}},
  \bibinfo {author} {\bibfnamefont {Y.}~\bibnamefont {Le~Coq}}, \bibinfo
  {author} {\bibfnamefont {M.}~\bibnamefont {Zawada}}, \bibinfo {author}
  {\bibfnamefont {J.}~\bibnamefont {Gu{\'e}na}}, \bibinfo {author}
  {\bibfnamefont {M.}~\bibnamefont {Abgrall}}, \bibinfo {author} {\bibfnamefont
  {M.}~\bibnamefont {Gurov}}, \bibinfo {author} {\bibfnamefont
  {P.}~\bibnamefont {Rosenbusch}}, \bibinfo {author} {\bibfnamefont {D.~G.}\
  \bibnamefont {Rovera}}, \bibinfo {author} {\bibfnamefont {B.}~\bibnamefont
  {Nag{\'o}rny}}, \bibinfo {author} {\bibfnamefont {R.}~\bibnamefont
  {Gartman}}, \bibinfo {author} {\bibfnamefont {P.~G.}\ \bibnamefont
  {Westergaard}}, \bibinfo {author} {\bibfnamefont {M.~E.}\ \bibnamefont
  {Tobar}}, \bibinfo {author} {\bibfnamefont {M.}~\bibnamefont {Lours}},
  \bibinfo {author} {\bibfnamefont {G.}~\bibnamefont {Santarelli}}, \bibinfo
  {author} {\bibfnamefont {A.}~\bibnamefont {Clairon}}, \bibinfo {author}
  {\bibfnamefont {S.}~\bibnamefont {Bize}}, \bibinfo {author} {\bibfnamefont
  {P.}~\bibnamefont {Laurent}}, \bibinfo {author} {\bibfnamefont
  {P.}~\bibnamefont {Lemonde}},\ and\ \bibinfo {author} {\bibfnamefont
  {J.}~\bibnamefont {Lodewyck}},\ }\bibfield  {title} {\bibinfo {title}
  {Experimental realization of an optical second with strontium lattice
  clocks},\ }\href {https://doi.org/10.1038/ncomms3109} {\bibfield  {journal}
  {\bibinfo  {journal} {Nature Communications}\ }\textbf {\bibinfo {volume}
  {4}},\ \bibinfo {pages} {2109} (\bibinfo {year} {2013})}\BibitemShut
  {NoStop}%
\bibitem [{\citenamefont {{Boulder Atomic Clock Optical Network (BACON)
  Collaboration*}}(2021)}]{boulderatomicclockopticalnetworkbaconcollaboration*FrequencyRatioMeasurements2021}%
  \BibitemOpen
  \bibfield  {author} {\bibinfo {author} {\bibnamefont {{Boulder Atomic Clock
  Optical Network (BACON) Collaboration*}}},\ }\bibfield  {title} {\bibinfo
  {title} {Frequency ratio measurements at 18-digit accuracy using an optical
  clock network},\ }\href {https://doi.org/10.1038/s41586-021-03253-4}
  {\bibfield  {journal} {\bibinfo  {journal} {Nature}\ }\textbf {\bibinfo
  {volume} {591}},\ \bibinfo {pages} {564} (\bibinfo {year}
  {2021})}\BibitemShut {NoStop}%
\bibitem [{\citenamefont {Hobson}\ \emph {et~al.}(2020)\citenamefont {Hobson},
  \citenamefont {Bowden}, \citenamefont {Vianello}, \citenamefont {Silva},
  \citenamefont {Baynham}, \citenamefont {Margolis}, \citenamefont {Baird},
  \citenamefont {Gill},\ and\ \citenamefont
  {Hill}}]{hobsonStrontiumOpticalLattice2020}%
  \BibitemOpen
  \bibfield  {author} {\bibinfo {author} {\bibfnamefont {R.}~\bibnamefont
  {Hobson}}, \bibinfo {author} {\bibfnamefont {W.}~\bibnamefont {Bowden}},
  \bibinfo {author} {\bibfnamefont {A.}~\bibnamefont {Vianello}}, \bibinfo
  {author} {\bibfnamefont {A.}~\bibnamefont {Silva}}, \bibinfo {author}
  {\bibfnamefont {C.~F.~A.}\ \bibnamefont {Baynham}}, \bibinfo {author}
  {\bibfnamefont {H.~S.}\ \bibnamefont {Margolis}}, \bibinfo {author}
  {\bibfnamefont {P.~E.~G.}\ \bibnamefont {Baird}}, \bibinfo {author}
  {\bibfnamefont {P.}~\bibnamefont {Gill}},\ and\ \bibinfo {author}
  {\bibfnamefont {I.~R.}\ \bibnamefont {Hill}},\ }\bibfield  {title} {\bibinfo
  {title} {A strontium optical lattice clock with 1 \texttimes{} 10
  {\textsuperscript{-17}} uncertainty and measurement of its absolute
  frequency},\ }\href {https://doi.org/10.1088/1681-7575/abb530} {\bibfield
  {journal} {\bibinfo  {journal} {Metrologia}\ }\textbf {\bibinfo {volume}
  {57}},\ \bibinfo {pages} {065026} (\bibinfo {year} {2020})}\BibitemShut
  {NoStop}%
\bibitem [{\citenamefont {Nemitz}\ \emph {et~al.}(2016)\citenamefont {Nemitz},
  \citenamefont {Ohkubo}, \citenamefont {Takamoto}, \citenamefont {Ushijima},
  \citenamefont {Das}, \citenamefont {Ohmae},\ and\ \citenamefont
  {Katori}}]{nemitzFrequencyRatioYb2016}%
  \BibitemOpen
  \bibfield  {author} {\bibinfo {author} {\bibfnamefont {N.}~\bibnamefont
  {Nemitz}}, \bibinfo {author} {\bibfnamefont {T.}~\bibnamefont {Ohkubo}},
  \bibinfo {author} {\bibfnamefont {M.}~\bibnamefont {Takamoto}}, \bibinfo
  {author} {\bibfnamefont {I.}~\bibnamefont {Ushijima}}, \bibinfo {author}
  {\bibfnamefont {M.}~\bibnamefont {Das}}, \bibinfo {author} {\bibfnamefont
  {N.}~\bibnamefont {Ohmae}},\ and\ \bibinfo {author} {\bibfnamefont
  {H.}~\bibnamefont {Katori}},\ }\bibfield  {title} {\bibinfo {title}
  {Frequency ratio of {{Yb}} and {{Sr}} clocks with 5 \texttimes{}
  10{\textsuperscript{-17}} uncertainty at 150 seconds averaging time},\ }\href
  {https://doi.org/10.1038/nphoton.2016.20} {\bibfield  {journal} {\bibinfo
  {journal} {Nature Photonics}\ }\textbf {\bibinfo {volume} {10}},\ \bibinfo
  {pages} {258} (\bibinfo {year} {2016})}\BibitemShut {NoStop}%
\bibitem [{\citenamefont {Ushijima}\ \emph {et~al.}(2015)\citenamefont
  {Ushijima}, \citenamefont {Takamoto}, \citenamefont {Das}, \citenamefont
  {Ohkubo},\ and\ \citenamefont
  {Katori}}]{ushijimaCryogenicOpticalLattice2015}%
  \BibitemOpen
  \bibfield  {author} {\bibinfo {author} {\bibfnamefont {I.}~\bibnamefont
  {Ushijima}}, \bibinfo {author} {\bibfnamefont {M.}~\bibnamefont {Takamoto}},
  \bibinfo {author} {\bibfnamefont {M.}~\bibnamefont {Das}}, \bibinfo {author}
  {\bibfnamefont {T.}~\bibnamefont {Ohkubo}},\ and\ \bibinfo {author}
  {\bibfnamefont {H.}~\bibnamefont {Katori}},\ }\bibfield  {title} {\bibinfo
  {title} {Cryogenic optical lattice clocks},\ }\href
  {https://doi.org/10.1038/nphoton.2015.5} {\bibfield  {journal} {\bibinfo
  {journal} {Nature Photonics}\ }\textbf {\bibinfo {volume} {9}},\ \bibinfo
  {pages} {185} (\bibinfo {year} {2015})}\BibitemShut {NoStop}%
\bibitem [{\citenamefont {Schwarz}\ \emph {et~al.}(2020)\citenamefont
  {Schwarz}, \citenamefont {D{\"o}rscher}, \citenamefont {{Al-Masoudi}},
  \citenamefont {Benkler}, \citenamefont {Legero}, \citenamefont {Sterr},
  \citenamefont {Weyers}, \citenamefont {Rahm}, \citenamefont {Lipphardt},\
  and\ \citenamefont {Lisdat}}]{schwarzLongTermMeasurement2020}%
  \BibitemOpen
  \bibfield  {author} {\bibinfo {author} {\bibfnamefont {R.}~\bibnamefont
  {Schwarz}}, \bibinfo {author} {\bibfnamefont {S.}~\bibnamefont
  {D{\"o}rscher}}, \bibinfo {author} {\bibfnamefont {A.}~\bibnamefont
  {{Al-Masoudi}}}, \bibinfo {author} {\bibfnamefont {E.}~\bibnamefont
  {Benkler}}, \bibinfo {author} {\bibfnamefont {T.}~\bibnamefont {Legero}},
  \bibinfo {author} {\bibfnamefont {U.}~\bibnamefont {Sterr}}, \bibinfo
  {author} {\bibfnamefont {S.}~\bibnamefont {Weyers}}, \bibinfo {author}
  {\bibfnamefont {J.}~\bibnamefont {Rahm}}, \bibinfo {author} {\bibfnamefont
  {B.}~\bibnamefont {Lipphardt}},\ and\ \bibinfo {author} {\bibfnamefont
  {C.}~\bibnamefont {Lisdat}},\ }\bibfield  {title} {\bibinfo {title} {Long
  term measurement of the {{Sr}} 87 clock frequency at the limit of primary
  {{Cs}} clocks},\ }\href {https://doi.org/10.1103/PhysRevResearch.2.033242}
  {\bibfield  {journal} {\bibinfo  {journal} {Physical Review Research}\
  }\textbf {\bibinfo {volume} {2}},\ \bibinfo {pages} {033242} (\bibinfo {year}
  {2020})}\BibitemShut {NoStop}%
\bibitem [{\citenamefont {Kim}\ \emph {et~al.}(2021)\citenamefont {Kim},
  \citenamefont {Heo}, \citenamefont {Park}, \citenamefont {Yu},\ and\
  \citenamefont {Lee}}]{kimAbsoluteFrequencyMeasurement2021a}%
  \BibitemOpen
  \bibfield  {author} {\bibinfo {author} {\bibfnamefont {H.}~\bibnamefont
  {Kim}}, \bibinfo {author} {\bibfnamefont {M.-S.}\ \bibnamefont {Heo}},
  \bibinfo {author} {\bibfnamefont {C.~Y.}\ \bibnamefont {Park}}, \bibinfo
  {author} {\bibfnamefont {D.-H.}\ \bibnamefont {Yu}},\ and\ \bibinfo {author}
  {\bibfnamefont {W.-K.}\ \bibnamefont {Lee}},\ }\bibfield  {title} {\bibinfo
  {title} {Absolute frequency measurement of the {\textsuperscript{171}} {{Yb}}
  optical lattice clock at {{KRISS}} using {{TAI}} for over a year},\ }\href
  {https://doi.org/10.1088/1681-7575/ac1950} {\bibfield  {journal} {\bibinfo
  {journal} {Metrologia}\ }\textbf {\bibinfo {volume} {58}},\ \bibinfo {pages}
  {055007} (\bibinfo {year} {2021})}\BibitemShut {NoStop}%
\bibitem [{\citenamefont {Lemonde}\ and\ \citenamefont
  {Wolf}(2005)}]{lemondeOpticalLatticeClock2005}%
  \BibitemOpen
  \bibfield  {author} {\bibinfo {author} {\bibfnamefont {P.}~\bibnamefont
  {Lemonde}}\ and\ \bibinfo {author} {\bibfnamefont {P.}~\bibnamefont {Wolf}},\
  }\bibfield  {title} {\bibinfo {title} {Optical lattice clock with atoms
  confined in a shallow trap},\ }\href
  {https://doi.org/10.1103/PhysRevA.72.033409} {\bibfield  {journal} {\bibinfo
  {journal} {Physical Review A}\ }\textbf {\bibinfo {volume} {72}},\ \bibinfo
  {pages} {033409} (\bibinfo {year} {2005})}\BibitemShut {NoStop}%
\bibitem [{\citenamefont {Aeppli}\ \emph {et~al.}(2022)\citenamefont {Aeppli},
  \citenamefont {Chu}, \citenamefont {Bothwell}, \citenamefont {Kennedy},
  \citenamefont {Kedar}, \citenamefont {He}, \citenamefont {Rey},\ and\
  \citenamefont {{Jun Ye}}}]{aeppliHamiltonianEngineeringSpinorbit2022}%
  \BibitemOpen
  \bibfield  {author} {\bibinfo {author} {\bibfnamefont {A.}~\bibnamefont
  {Aeppli}}, \bibinfo {author} {\bibfnamefont {A.}~\bibnamefont {Chu}},
  \bibinfo {author} {\bibfnamefont {T.}~\bibnamefont {Bothwell}}, \bibinfo
  {author} {\bibfnamefont {C.~J.}\ \bibnamefont {Kennedy}}, \bibinfo {author}
  {\bibfnamefont {D.}~\bibnamefont {Kedar}}, \bibinfo {author} {\bibfnamefont
  {P.}~\bibnamefont {He}}, \bibinfo {author} {\bibfnamefont {A.~M.}\
  \bibnamefont {Rey}},\ and\ \bibinfo {author} {\bibnamefont {{Jun Ye}}},\
  }\bibfield  {title} {\bibinfo {title} {Hamiltonian engineering of spin-orbit
  coupled fermions in a {{Wannier-Stark}} optical lattice clock},\ }\href
  {https://doi.org/10.1126/sciadv.adc9242} {\bibfield  {journal} {\bibinfo
  {journal} {Science Advances}\ }\textbf {\bibinfo {volume} {8}},\ \bibinfo
  {pages} {eadc9242} (\bibinfo {year} {2022})}\BibitemShut {NoStop}%
\bibitem [{\citenamefont {Ushijima}\ \emph {et~al.}(2018)\citenamefont
  {Ushijima}, \citenamefont {Takamoto},\ and\ \citenamefont
  {Katori}}]{ushijimaOperationalMagicIntensity2018}%
  \BibitemOpen
  \bibfield  {author} {\bibinfo {author} {\bibfnamefont {I.}~\bibnamefont
  {Ushijima}}, \bibinfo {author} {\bibfnamefont {M.}~\bibnamefont {Takamoto}},\
  and\ \bibinfo {author} {\bibfnamefont {H.}~\bibnamefont {Katori}},\
  }\bibfield  {title} {\bibinfo {title} {Operational {{Magic Intensity}} for
  {{Sr Optical Lattice Clocks}}},\ }\href
  {https://doi.org/10.1103/PhysRevLett.121.263202} {\bibfield  {journal}
  {\bibinfo  {journal} {Physical Review Letters}\ }\textbf {\bibinfo {volume}
  {121}},\ \bibinfo {pages} {263202} (\bibinfo {year} {2018})}\BibitemShut
  {NoStop}%
\bibitem [{\citenamefont {Katori}\ \emph {et~al.}(2015)\citenamefont {Katori},
  \citenamefont {Ovsiannikov}, \citenamefont {Marmo},\ and\ \citenamefont
  {Palchikov}}]{katoriStrategiesReducingLight2015}%
  \BibitemOpen
  \bibfield  {author} {\bibinfo {author} {\bibfnamefont {H.}~\bibnamefont
  {Katori}}, \bibinfo {author} {\bibfnamefont {V.~D.}\ \bibnamefont
  {Ovsiannikov}}, \bibinfo {author} {\bibfnamefont {S.~I.}\ \bibnamefont
  {Marmo}},\ and\ \bibinfo {author} {\bibfnamefont {V.~G.}\ \bibnamefont
  {Palchikov}},\ }\bibfield  {title} {\bibinfo {title} {Strategies for reducing
  the light shift in atomic clocks},\ }\href
  {https://doi.org/10.1103/PhysRevA.91.052503} {\bibfield  {journal} {\bibinfo
  {journal} {Physical Review A}\ }\textbf {\bibinfo {volume} {91}},\ \bibinfo
  {pages} {052503} (\bibinfo {year} {2015})}\BibitemShut {NoStop}%
\bibitem [{\citenamefont {Beloy}\ \emph {et~al.}(2020)\citenamefont {Beloy},
  \citenamefont {McGrew}, \citenamefont {Zhang}, \citenamefont {Nicolodi},
  \citenamefont {Fasano}, \citenamefont {Hassan}, \citenamefont {Brown},\ and\
  \citenamefont {Ludlow}}]{beloyModelingMotionalEnergy2020}%
  \BibitemOpen
  \bibfield  {author} {\bibinfo {author} {\bibfnamefont {K.}~\bibnamefont
  {Beloy}}, \bibinfo {author} {\bibfnamefont {W.~F.}\ \bibnamefont {McGrew}},
  \bibinfo {author} {\bibfnamefont {X.}~\bibnamefont {Zhang}}, \bibinfo
  {author} {\bibfnamefont {D.}~\bibnamefont {Nicolodi}}, \bibinfo {author}
  {\bibfnamefont {R.~J.}\ \bibnamefont {Fasano}}, \bibinfo {author}
  {\bibfnamefont {Y.~S.}\ \bibnamefont {Hassan}}, \bibinfo {author}
  {\bibfnamefont {R.~C.}\ \bibnamefont {Brown}},\ and\ \bibinfo {author}
  {\bibfnamefont {A.~D.}\ \bibnamefont {Ludlow}},\ }\bibfield  {title}
  {\bibinfo {title} {Modeling motional energy spectra and lattice light shifts
  in optical lattice clocks},\ }\href
  {https://doi.org/10.1103/PhysRevA.101.053416} {\bibfield  {journal} {\bibinfo
   {journal} {Physical Review A}\ }\textbf {\bibinfo {volume} {101}},\ \bibinfo
  {pages} {053416} (\bibinfo {year} {2020})}\BibitemShut {NoStop}%
\bibitem [{See()}]{SeeSupplementalMaterial}%
  \BibitemOpen
  \bibinfo {title} {See {{Supplemental Material}} at [{{URL}} will be inserted
  by publisher] for more information.}\BibitemShut {Stop}%
\bibitem [{\citenamefont {Blatt}\ \emph {et~al.}(2009)\citenamefont {Blatt},
  \citenamefont {Thomsen}, \citenamefont {Campbell}, \citenamefont {Ludlow},
  \citenamefont {Swallows}, \citenamefont {Martin}, \citenamefont {Boyd},\ and\
  \citenamefont {Ye}}]{blattRabiSpectroscopyExcitation2009}%
  \BibitemOpen
\bibfield  {title} {  }\bibfield  {author} {\bibinfo {author} {\bibfnamefont
  {S.}~\bibnamefont {Blatt}}, \bibinfo {author} {\bibfnamefont {J.~W.}\
  \bibnamefont {Thomsen}}, \bibinfo {author} {\bibfnamefont {G.~K.}\
  \bibnamefont {Campbell}}, \bibinfo {author} {\bibfnamefont {A.~D.}\
  \bibnamefont {Ludlow}}, \bibinfo {author} {\bibfnamefont {M.~D.}\
  \bibnamefont {Swallows}}, \bibinfo {author} {\bibfnamefont {M.~J.}\
  \bibnamefont {Martin}}, \bibinfo {author} {\bibfnamefont {M.~M.}\
  \bibnamefont {Boyd}},\ and\ \bibinfo {author} {\bibfnamefont
  {J.}~\bibnamefont {Ye}},\ }\bibfield  {title} {\bibinfo {title} {Rabi
  spectroscopy and excitation inhomogeneity in a one-dimensional optical
  lattice clock},\ }\href {https://doi.org/10.1103/PhysRevA.80.052703}
  {\bibfield  {journal} {\bibinfo  {journal} {Physical Review A}\ }\textbf
  {\bibinfo {volume} {80}},\ \bibinfo {pages} {052703} (\bibinfo {year}
  {2009})}\BibitemShut {NoStop}%
\bibitem [{\citenamefont {Oelker}\ \emph {et~al.}(2019)\citenamefont {Oelker},
  \citenamefont {Hutson}, \citenamefont {Kennedy}, \citenamefont {Sonderhouse},
  \citenamefont {Bothwell}, \citenamefont {Goban}, \citenamefont {Kedar},
  \citenamefont {Sanner}, \citenamefont {Robinson}, \citenamefont {Marti},
  \citenamefont {Matei}, \citenamefont {Legero}, \citenamefont {Giunta},
  \citenamefont {Holzwarth}, \citenamefont {Riehle}, \citenamefont {Sterr},\
  and\ \citenamefont {Ye}}]{oelkerDemonstration10172019}%
  \BibitemOpen
  \bibfield  {author} {\bibinfo {author} {\bibfnamefont {E.}~\bibnamefont
  {Oelker}}, \bibinfo {author} {\bibfnamefont {R.~B.}\ \bibnamefont {Hutson}},
  \bibinfo {author} {\bibfnamefont {C.~J.}\ \bibnamefont {Kennedy}}, \bibinfo
  {author} {\bibfnamefont {L.}~\bibnamefont {Sonderhouse}}, \bibinfo {author}
  {\bibfnamefont {T.}~\bibnamefont {Bothwell}}, \bibinfo {author}
  {\bibfnamefont {A.}~\bibnamefont {Goban}}, \bibinfo {author} {\bibfnamefont
  {D.}~\bibnamefont {Kedar}}, \bibinfo {author} {\bibfnamefont
  {C.}~\bibnamefont {Sanner}}, \bibinfo {author} {\bibfnamefont {J.~M.}\
  \bibnamefont {Robinson}}, \bibinfo {author} {\bibfnamefont {G.~E.}\
  \bibnamefont {Marti}}, \bibinfo {author} {\bibfnamefont {D.~G.}\ \bibnamefont
  {Matei}}, \bibinfo {author} {\bibfnamefont {T.}~\bibnamefont {Legero}},
  \bibinfo {author} {\bibfnamefont {M.}~\bibnamefont {Giunta}}, \bibinfo
  {author} {\bibfnamefont {R.}~\bibnamefont {Holzwarth}}, \bibinfo {author}
  {\bibfnamefont {F.}~\bibnamefont {Riehle}}, \bibinfo {author} {\bibfnamefont
  {U.}~\bibnamefont {Sterr}},\ and\ \bibinfo {author} {\bibfnamefont
  {J.}~\bibnamefont {Ye}},\ }\bibfield  {title} {\bibinfo {title}
  {Demonstration of 4.8 \texttimes{} 10{\textsuperscript{-17}} stability at 1 s
  for two independent optical clocks},\ }\href
  {https://doi.org/10.1038/s41566-019-0493-4} {\bibfield  {journal} {\bibinfo
  {journal} {Nature Photonics}\ }\textbf {\bibinfo {volume} {13}},\ \bibinfo
  {pages} {714} (\bibinfo {year} {2019})}\BibitemShut {NoStop}%
\bibitem [{\citenamefont {Matei}\ \emph {et~al.}(2017)\citenamefont {Matei},
  \citenamefont {Legero}, \citenamefont {H{\"a}fner}, \citenamefont {Grebing},
  \citenamefont {Weyrich}, \citenamefont {Zhang}, \citenamefont {Sonderhouse},
  \citenamefont {Robinson}, \citenamefont {Ye}, \citenamefont {Riehle},\ and\
  \citenamefont {Sterr}}]{mateiMmLasersSub102017}%
  \BibitemOpen
  \bibfield  {author} {\bibinfo {author} {\bibfnamefont {D.~G.}\ \bibnamefont
  {Matei}}, \bibinfo {author} {\bibfnamefont {T.}~\bibnamefont {Legero}},
  \bibinfo {author} {\bibfnamefont {S.}~\bibnamefont {H{\"a}fner}}, \bibinfo
  {author} {\bibfnamefont {C.}~\bibnamefont {Grebing}}, \bibinfo {author}
  {\bibfnamefont {R.}~\bibnamefont {Weyrich}}, \bibinfo {author} {\bibfnamefont
  {W.}~\bibnamefont {Zhang}}, \bibinfo {author} {\bibfnamefont
  {L.}~\bibnamefont {Sonderhouse}}, \bibinfo {author} {\bibfnamefont {J.~M.}\
  \bibnamefont {Robinson}}, \bibinfo {author} {\bibfnamefont {J.}~\bibnamefont
  {Ye}}, \bibinfo {author} {\bibfnamefont {F.}~\bibnamefont {Riehle}},\ and\
  \bibinfo {author} {\bibfnamefont {U.}~\bibnamefont {Sterr}},\ }\bibfield
  {title} {\bibinfo {title} {1.5 {$M$}m {{Lasers}} with {{Sub-10 mHz
  Linewidth}}},\ }\href {https://doi.org/10.1103/PhysRevLett.118.263202}
  {\bibfield  {journal} {\bibinfo  {journal} {Physical Review Letters}\
  }\textbf {\bibinfo {volume} {118}},\ \bibinfo {pages} {263202} (\bibinfo
  {year} {2017})}\BibitemShut {NoStop}%
\bibitem [{\citenamefont {Nicholson}\ \emph {et~al.}(2012)\citenamefont
  {Nicholson}, \citenamefont {Martin}, \citenamefont {Williams}, \citenamefont
  {Bloom}, \citenamefont {Bishof}, \citenamefont {Swallows}, \citenamefont
  {Campbell},\ and\ \citenamefont
  {Ye}}]{nicholsonComparisonTwoIndependent2012}%
  \BibitemOpen
  \bibfield  {author} {\bibinfo {author} {\bibfnamefont {T.~L.}\ \bibnamefont
  {Nicholson}}, \bibinfo {author} {\bibfnamefont {M.~J.}\ \bibnamefont
  {Martin}}, \bibinfo {author} {\bibfnamefont {J.~R.}\ \bibnamefont
  {Williams}}, \bibinfo {author} {\bibfnamefont {B.~J.}\ \bibnamefont {Bloom}},
  \bibinfo {author} {\bibfnamefont {M.}~\bibnamefont {Bishof}}, \bibinfo
  {author} {\bibfnamefont {M.~D.}\ \bibnamefont {Swallows}}, \bibinfo {author}
  {\bibfnamefont {S.~L.}\ \bibnamefont {Campbell}},\ and\ \bibinfo {author}
  {\bibfnamefont {J.}~\bibnamefont {Ye}},\ }\bibfield  {title} {\bibinfo
  {title} {Comparison of {{Two Independent Sr Optical Clocks}} with 1
  \texttimes{} 10 {\textsuperscript{-17}} {{Stability}} at
  10{\textsuperscript{3}} s},\ }\href
  {https://doi.org/10.1103/PhysRevLett.109.230801} {\bibfield  {journal}
  {\bibinfo  {journal} {Physical Review Letters}\ }\textbf {\bibinfo {volume}
  {109}},\ \bibinfo {pages} {230801} (\bibinfo {year} {2012})}\BibitemShut
  {NoStop}%
\bibitem [{\citenamefont {Westergaard}\ \emph {et~al.}(2011)\citenamefont
  {Westergaard}, \citenamefont {Lodewyck}, \citenamefont {Lorini},
  \citenamefont {Lecallier}, \citenamefont {Burt}, \citenamefont {Zawada},
  \citenamefont {Millo},\ and\ \citenamefont
  {Lemonde}}]{westergaardLatticeInducedFrequencyShifts2011}%
  \BibitemOpen
  \bibfield  {author} {\bibinfo {author} {\bibfnamefont {P.~G.}\ \bibnamefont
  {Westergaard}}, \bibinfo {author} {\bibfnamefont {J.}~\bibnamefont
  {Lodewyck}}, \bibinfo {author} {\bibfnamefont {L.}~\bibnamefont {Lorini}},
  \bibinfo {author} {\bibfnamefont {A.}~\bibnamefont {Lecallier}}, \bibinfo
  {author} {\bibfnamefont {E.~A.}\ \bibnamefont {Burt}}, \bibinfo {author}
  {\bibfnamefont {M.}~\bibnamefont {Zawada}}, \bibinfo {author} {\bibfnamefont
  {J.}~\bibnamefont {Millo}},\ and\ \bibinfo {author} {\bibfnamefont
  {P.}~\bibnamefont {Lemonde}},\ }\bibfield  {title} {\bibinfo {title}
  {Lattice-{{Induced Frequency Shifts}} in {{Sr Optical Lattice Clocks}} at the
  10{\textsuperscript{-17}} {{Level}}},\ }\href
  {https://doi.org/10.1103/PhysRevLett.106.210801} {\bibfield  {journal}
  {\bibinfo  {journal} {Physical Review Letters}\ }\textbf {\bibinfo {volume}
  {106}},\ \bibinfo {pages} {210801} (\bibinfo {year} {2011})}\BibitemShut
  {NoStop}%
\bibitem [{\citenamefont {Shi}\ \emph {et~al.}(2015)\citenamefont {Shi},
  \citenamefont {Robyr}, \citenamefont {Eismann}, \citenamefont {Zawada},
  \citenamefont {Lorini}, \citenamefont {Le~Targat},\ and\ \citenamefont
  {Lodewyck}}]{shiPolarizabilitiesSr872015}%
  \BibitemOpen
  \bibfield  {author} {\bibinfo {author} {\bibfnamefont {C.}~\bibnamefont
  {Shi}}, \bibinfo {author} {\bibfnamefont {J.-L.}\ \bibnamefont {Robyr}},
  \bibinfo {author} {\bibfnamefont {U.}~\bibnamefont {Eismann}}, \bibinfo
  {author} {\bibfnamefont {M.}~\bibnamefont {Zawada}}, \bibinfo {author}
  {\bibfnamefont {L.}~\bibnamefont {Lorini}}, \bibinfo {author} {\bibfnamefont
  {R.}~\bibnamefont {Le~Targat}},\ and\ \bibinfo {author} {\bibfnamefont
  {J.}~\bibnamefont {Lodewyck}},\ }\bibfield  {title} {\bibinfo {title}
  {Polarizabilities of the {{Sr}} 87 clock transition},\ }\href
  {https://doi.org/10.1103/PhysRevA.92.012516} {\bibfield  {journal} {\bibinfo
  {journal} {Physical Review A}\ }\textbf {\bibinfo {volume} {92}},\ \bibinfo
  {pages} {012516} (\bibinfo {year} {2015})}\BibitemShut {NoStop}%
\bibitem [{\citenamefont {Fasano}\ \emph {et~al.}(2021)\citenamefont {Fasano},
  \citenamefont {Chen}, \citenamefont {McGrew}, \citenamefont {Brand},
  \citenamefont {Fox},\ and\ \citenamefont
  {Ludlow}}]{fasanoCharacterizationSuppressionBackground2021}%
  \BibitemOpen
  \bibfield  {author} {\bibinfo {author} {\bibfnamefont {R.}~\bibnamefont
  {Fasano}}, \bibinfo {author} {\bibfnamefont {Y.}~\bibnamefont {Chen}},
  \bibinfo {author} {\bibfnamefont {W.}~\bibnamefont {McGrew}}, \bibinfo
  {author} {\bibfnamefont {W.}~\bibnamefont {Brand}}, \bibinfo {author}
  {\bibfnamefont {R.}~\bibnamefont {Fox}},\ and\ \bibinfo {author}
  {\bibfnamefont {A.}~\bibnamefont {Ludlow}},\ }\bibfield  {title} {\bibinfo
  {title} {Characterization and {{Suppression}} of {{Background Light Shifts}}
  in an {{Optical Lattice Clock}}},\ }\href
  {https://doi.org/10.1103/PhysRevApplied.15.044016} {\bibfield  {journal}
  {\bibinfo  {journal} {Physical Review Applied}\ }\textbf {\bibinfo {volume}
  {15}},\ \bibinfo {pages} {044016} (\bibinfo {year} {2021})}\BibitemShut
  {NoStop}%
\bibitem [{\citenamefont {D{\"o}rscher}\ \emph {et~al.}(2022)\citenamefont
  {D{\"o}rscher}, \citenamefont {Klose}, \citenamefont {Palli},\ and\
  \citenamefont {Lisdat}}]{dorscherExperimentalDeterminationE2M12022}%
  \BibitemOpen
  \bibfield  {author} {\bibinfo {author} {\bibfnamefont {S.}~\bibnamefont
  {D{\"o}rscher}}, \bibinfo {author} {\bibfnamefont {J.}~\bibnamefont {Klose}},
  \bibinfo {author} {\bibfnamefont {S.~M.}\ \bibnamefont {Palli}},\ and\
  \bibinfo {author} {\bibfnamefont {C.}~\bibnamefont {Lisdat}},\ }\href
  {http://arxiv.org/abs/2210.14727} {\bibinfo {title} {Experimental
  determination of the {{E2-M1}} polarizability of the strontium clock
  transition}} (\bibinfo {year} {2022}),\ \Eprint
  {https://arxiv.org/abs/2210.14727} {arXiv:2210.14727 [physics]} \BibitemShut
  {NoStop}%
\bibitem [{\citenamefont {Porsev}\ \emph {et~al.}(2018)\citenamefont {Porsev},
  \citenamefont {Safronova}, \citenamefont {Safronova},\ and\ \citenamefont
  {Kozlov}}]{porsevMultipolarPolarizabilitiesHyperpolarizabilities2018}%
  \BibitemOpen
  \bibfield  {author} {\bibinfo {author} {\bibfnamefont {S.~G.}\ \bibnamefont
  {Porsev}}, \bibinfo {author} {\bibfnamefont {M.~S.}\ \bibnamefont
  {Safronova}}, \bibinfo {author} {\bibfnamefont {U.~I.}\ \bibnamefont
  {Safronova}},\ and\ \bibinfo {author} {\bibfnamefont {M.~G.}\ \bibnamefont
  {Kozlov}},\ }\bibfield  {title} {\bibinfo {title} {Multipolar
  {{Polarizabilities}} and {{Hyperpolarizabilities}} in the {{Sr Optical
  Lattice Clock}}},\ }\href {https://doi.org/10.1103/PhysRevLett.120.063204}
  {\bibfield  {journal} {\bibinfo  {journal} {Physical Review Letters}\
  }\textbf {\bibinfo {volume} {120}},\ \bibinfo {pages} {063204} (\bibinfo
  {year} {2018})}\BibitemShut {NoStop}%
\bibitem [{\citenamefont {Wu}\ \emph {et~al.}(2019)\citenamefont {Wu},
  \citenamefont {Tang}, \citenamefont {Shi},\ and\ \citenamefont
  {Tang}}]{wuDynamicMultipolarPolarizabilities2019}%
  \BibitemOpen
  \bibfield  {author} {\bibinfo {author} {\bibfnamefont {F.-F.}\ \bibnamefont
  {Wu}}, \bibinfo {author} {\bibfnamefont {Y.-B.}\ \bibnamefont {Tang}},
  \bibinfo {author} {\bibfnamefont {T.-Y.}\ \bibnamefont {Shi}},\ and\ \bibinfo
  {author} {\bibfnamefont {L.-Y.}\ \bibnamefont {Tang}},\ }\bibfield  {title}
  {\bibinfo {title} {Dynamic multipolar polarizabilities and
  hyperpolarizabilities of the {{Sr}} lattice clock},\ }\href
  {https://doi.org/10.1103/PhysRevA.100.042514} {\bibfield  {journal} {\bibinfo
   {journal} {Physical Review A}\ }\textbf {\bibinfo {volume} {100}},\ \bibinfo
  {pages} {042514} (\bibinfo {year} {2019})}\BibitemShut {NoStop}%
\bibitem [{\citenamefont {Ovsiannikov}\ \emph {et~al.}(2016)\citenamefont
  {Ovsiannikov}, \citenamefont {Marmo}, \citenamefont {Palchikov},\ and\
  \citenamefont {Katori}}]{ovsiannikovHigherorderEffectsPrecision2016}%
  \BibitemOpen
  \bibfield  {author} {\bibinfo {author} {\bibfnamefont {V.~D.}\ \bibnamefont
  {Ovsiannikov}}, \bibinfo {author} {\bibfnamefont {S.~I.}\ \bibnamefont
  {Marmo}}, \bibinfo {author} {\bibfnamefont {V.~G.}\ \bibnamefont
  {Palchikov}},\ and\ \bibinfo {author} {\bibfnamefont {H.}~\bibnamefont
  {Katori}},\ }\bibfield  {title} {\bibinfo {title} {Higher-order effects on
  the precision of clocks of neutral atoms in optical lattices},\ }\href
  {https://doi.org/10.1103/PhysRevA.93.043420} {\bibfield  {journal} {\bibinfo
  {journal} {Physical Review A}\ }\textbf {\bibinfo {volume} {93}},\ \bibinfo
  {pages} {043420} (\bibinfo {year} {2016})}\BibitemShut {NoStop}%
\bibitem [{\citenamefont {Ovsiannikov}\ \emph {et~al.}(2013)\citenamefont
  {Ovsiannikov}, \citenamefont {Pal'chikov}, \citenamefont {Taichenachev},
  \citenamefont {Yudin},\ and\ \citenamefont
  {Katori}}]{ovsiannikovMultipoleNonlinearAnharmonic2013}%
  \BibitemOpen
  \bibfield  {author} {\bibinfo {author} {\bibfnamefont {V.~D.}\ \bibnamefont
  {Ovsiannikov}}, \bibinfo {author} {\bibfnamefont {V.~G.}\ \bibnamefont
  {Pal'chikov}}, \bibinfo {author} {\bibfnamefont {A.~V.}\ \bibnamefont
  {Taichenachev}}, \bibinfo {author} {\bibfnamefont {V.~I.}\ \bibnamefont
  {Yudin}},\ and\ \bibinfo {author} {\bibfnamefont {H.}~\bibnamefont
  {Katori}},\ }\bibfield  {title} {\bibinfo {title} {Multipole, nonlinear, and
  anharmonic uncertainties of clocks of {{Sr}} atoms in an optical lattice},\
  }\href {https://doi.org/10.1103/PhysRevA.88.013405} {\bibfield  {journal}
  {\bibinfo  {journal} {Physical Review A}\ }\textbf {\bibinfo {volume} {88}},\
  \bibinfo {pages} {013405} (\bibinfo {year} {2013})}\BibitemShut {NoStop}%
\bibitem [{\citenamefont {Wu}\ \emph {et~al.}(2023)\citenamefont {Wu},
  \citenamefont {Shi},\ and\ \citenamefont
  {Tang}}]{wuContributionsNegativeenergyStates2023}%
  \BibitemOpen
  \bibfield  {author} {\bibinfo {author} {\bibfnamefont {F.-F.}\ \bibnamefont
  {Wu}}, \bibinfo {author} {\bibfnamefont {T.-Y.}\ \bibnamefont {Shi}},\ and\
  \bibinfo {author} {\bibfnamefont {L.-Y.}\ \bibnamefont {Tang}},\ }\href
  {http://arxiv.org/abs/2301.06740} {\bibinfo {title} {Contributions of
  negative-energy states to the {{E2-M1}} polarizability of the {{Sr}} clock}}
  (\bibinfo {year} {2023}),\ \Eprint {https://arxiv.org/abs/2301.06740}
  {arXiv:2301.06740 [physics]} \BibitemShut {NoStop}%
\bibitem [{\citenamefont {Nicholson}\ \emph {et~al.}(2015)\citenamefont
  {Nicholson}, \citenamefont {Campbell}, \citenamefont {Hutson}, \citenamefont
  {Marti}, \citenamefont {Bloom}, \citenamefont {McNally}, \citenamefont
  {Zhang}, \citenamefont {Barrett}, \citenamefont {Safronova}, \citenamefont
  {Strouse}, \citenamefont {Tew},\ and\ \citenamefont
  {Ye}}]{nicholsonSystematicEvaluationAtomic2015}%
  \BibitemOpen
  \bibfield  {author} {\bibinfo {author} {\bibfnamefont {T.}~\bibnamefont
  {Nicholson}}, \bibinfo {author} {\bibfnamefont {S.}~\bibnamefont {Campbell}},
  \bibinfo {author} {\bibfnamefont {R.}~\bibnamefont {Hutson}}, \bibinfo
  {author} {\bibfnamefont {G.}~\bibnamefont {Marti}}, \bibinfo {author}
  {\bibfnamefont {B.}~\bibnamefont {Bloom}}, \bibinfo {author} {\bibfnamefont
  {R.}~\bibnamefont {McNally}}, \bibinfo {author} {\bibfnamefont
  {W.}~\bibnamefont {Zhang}}, \bibinfo {author} {\bibfnamefont
  {M.}~\bibnamefont {Barrett}}, \bibinfo {author} {\bibfnamefont
  {M.}~\bibnamefont {Safronova}}, \bibinfo {author} {\bibfnamefont
  {G.}~\bibnamefont {Strouse}}, \bibinfo {author} {\bibfnamefont
  {W.}~\bibnamefont {Tew}},\ and\ \bibinfo {author} {\bibfnamefont
  {J.}~\bibnamefont {Ye}},\ }\bibfield  {title} {\bibinfo {title} {Systematic
  evaluation of an atomic clock at 2 \texttimes{} 10{\textsuperscript{-18}}
  total uncertainty},\ }\href {https://doi.org/10.1038/ncomms7896} {\bibfield
  {journal} {\bibinfo  {journal} {Nature Communications}\ }\textbf {\bibinfo
  {volume} {6}},\ \bibinfo {pages} {6896} (\bibinfo {year} {2015})}\BibitemShut
  {NoStop}%
\end{thebibliography}%


\begin{thebibliography}{13}%
\makeatletter
\providecommand \@ifxundefined [1]{%
 \@ifx{#1\undefined}
}%
\providecommand \@ifnum [1]{%
 \ifnum #1\expandafter \@firstoftwo
 \else \expandafter \@secondoftwo
 \fi
}%
\providecommand \@ifx [1]{%
 \ifx #1\expandafter \@firstoftwo
 \else \expandafter \@secondoftwo
 \fi
}%
\providecommand \natexlab [1]{#1}%
\providecommand \enquote  [1]{``#1''}%
\providecommand \bibnamefont  [1]{#1}%
\providecommand \bibfnamefont [1]{#1}%
\providecommand \citenamefont [1]{#1}%
\providecommand \href@noop [0]{\@secondoftwo}%
\providecommand \href [0]{\begingroup \@sanitize@url \@href}%
\providecommand \@href[1]{\@@startlink{#1}\@@href}%
\providecommand \@@href[1]{\endgroup#1\@@endlink}%
\providecommand \@sanitize@url [0]{\catcode `\\12\catcode `\$12\catcode
  `\&12\catcode `\#12\catcode `\^12\catcode `\_12\catcode `\%12\relax}%
\providecommand \@@startlink[1]{}%
\providecommand \@@endlink[0]{}%
\providecommand \url  [0]{\begingroup\@sanitize@url \@url }%
\providecommand \@url [1]{\endgroup\@href {#1}{\urlprefix }}%
\providecommand \urlprefix  [0]{URL }%
\providecommand \Eprint [0]{\href }%
\providecommand \doibase [0]{https://doi.org/}%
\providecommand \selectlanguage [0]{\@gobble}%
\providecommand \bibinfo  [0]{\@secondoftwo}%
\providecommand \bibfield  [0]{\@secondoftwo}%
\providecommand \translation [1]{[#1]}%
\providecommand \BibitemOpen [0]{}%
\providecommand \bibitemStop [0]{}%
\providecommand \bibitemNoStop [0]{.\EOS\space}%
\providecommand \EOS [0]{\spacefactor3000\relax}%
\providecommand \BibitemShut  [1]{\csname bibitem#1\endcsname}%
\let\auto@bib@innerbib\@empty
\bibitem [{\citenamefont {Blatt}\ \emph {et~al.}(2009)\citenamefont {Blatt},
  \citenamefont {Thomsen}, \citenamefont {Campbell}, \citenamefont {Ludlow},
  \citenamefont {Swallows}, \citenamefont {Martin}, \citenamefont {Boyd},\ and\
  \citenamefont {Ye}}]{blattRabiSpectroscopyExcitation2009}%
  \BibitemOpen
  \bibfield  {author} {\bibinfo {author} {\bibfnamefont {S.}~\bibnamefont
  {Blatt}}, \bibinfo {author} {\bibfnamefont {J.~W.}\ \bibnamefont {Thomsen}},
  \bibinfo {author} {\bibfnamefont {G.~K.}\ \bibnamefont {Campbell}}, \bibinfo
  {author} {\bibfnamefont {A.~D.}\ \bibnamefont {Ludlow}}, \bibinfo {author}
  {\bibfnamefont {M.~D.}\ \bibnamefont {Swallows}}, \bibinfo {author}
  {\bibfnamefont {M.~J.}\ \bibnamefont {Martin}}, \bibinfo {author}
  {\bibfnamefont {M.~M.}\ \bibnamefont {Boyd}},\ and\ \bibinfo {author}
  {\bibfnamefont {J.}~\bibnamefont {Ye}},\ }\bibfield  {title} {\bibinfo
  {title} {Rabi spectroscopy and excitation inhomogeneity in a one-dimensional
  optical lattice clock},\ }\href {https://doi.org/10.1103/PhysRevA.80.052703}
  {\bibfield  {journal} {\bibinfo  {journal} {Physical Review A}\ }\textbf
  {\bibinfo {volume} {80}},\ \bibinfo {pages} {052703} (\bibinfo {year}
  {2009})}\BibitemShut {NoStop}%
\bibitem [{\citenamefont {Katori}\ \emph {et~al.}(2015)\citenamefont {Katori},
  \citenamefont {Ovsiannikov}, \citenamefont {Marmo},\ and\ \citenamefont
  {Palchikov}}]{katoriStrategiesReducingLight2015}%
  \BibitemOpen
  \bibfield  {author} {\bibinfo {author} {\bibfnamefont {H.}~\bibnamefont
  {Katori}}, \bibinfo {author} {\bibfnamefont {V.~D.}\ \bibnamefont
  {Ovsiannikov}}, \bibinfo {author} {\bibfnamefont {S.~I.}\ \bibnamefont
  {Marmo}},\ and\ \bibinfo {author} {\bibfnamefont {V.~G.}\ \bibnamefont
  {Palchikov}},\ }\bibfield  {title} {\bibinfo {title} {Strategies for reducing
  the light shift in atomic clocks},\ }\href
  {https://doi.org/10.1103/PhysRevA.91.052503} {\bibfield  {journal} {\bibinfo
  {journal} {Physical Review A}\ }\textbf {\bibinfo {volume} {91}},\ \bibinfo
  {pages} {052503} (\bibinfo {year} {2015})}\BibitemShut {NoStop}%
\bibitem [{\citenamefont {Ushijima}\ \emph {et~al.}(2018)\citenamefont
  {Ushijima}, \citenamefont {Takamoto},\ and\ \citenamefont
  {Katori}}]{ushijimaOperationalMagicIntensity2018}%
  \BibitemOpen
  \bibfield  {author} {\bibinfo {author} {\bibfnamefont {I.}~\bibnamefont
  {Ushijima}}, \bibinfo {author} {\bibfnamefont {M.}~\bibnamefont {Takamoto}},\
  and\ \bibinfo {author} {\bibfnamefont {H.}~\bibnamefont {Katori}},\
  }\bibfield  {title} {\bibinfo {title} {Operational {{Magic Intensity}} for
  {{Sr Optical Lattice Clocks}}},\ }\href
  {https://doi.org/10.1103/PhysRevLett.121.263202} {\bibfield  {journal}
  {\bibinfo  {journal} {Physical Review Letters}\ }\textbf {\bibinfo {volume}
  {121}},\ \bibinfo {pages} {263202} (\bibinfo {year} {2018})}\BibitemShut
  {NoStop}%
\bibitem [{\citenamefont {Beloy}\ \emph {et~al.}(2020)\citenamefont {Beloy},
  \citenamefont {McGrew}, \citenamefont {Zhang}, \citenamefont {Nicolodi},
  \citenamefont {Fasano}, \citenamefont {Hassan}, \citenamefont {Brown},\ and\
  \citenamefont {Ludlow}}]{beloyModelingMotionalEnergy2020}%
  \BibitemOpen
  \bibfield  {author} {\bibinfo {author} {\bibfnamefont {K.}~\bibnamefont
  {Beloy}}, \bibinfo {author} {\bibfnamefont {W.~F.}\ \bibnamefont {McGrew}},
  \bibinfo {author} {\bibfnamefont {X.}~\bibnamefont {Zhang}}, \bibinfo
  {author} {\bibfnamefont {D.}~\bibnamefont {Nicolodi}}, \bibinfo {author}
  {\bibfnamefont {R.~J.}\ \bibnamefont {Fasano}}, \bibinfo {author}
  {\bibfnamefont {Y.~S.}\ \bibnamefont {Hassan}}, \bibinfo {author}
  {\bibfnamefont {R.~C.}\ \bibnamefont {Brown}},\ and\ \bibinfo {author}
  {\bibfnamefont {A.~D.}\ \bibnamefont {Ludlow}},\ }\bibfield  {title}
  {\bibinfo {title} {Modeling motional energy spectra and lattice light shifts
  in optical lattice clocks},\ }\href
  {https://doi.org/10.1103/PhysRevA.101.053416} {\bibfield  {journal} {\bibinfo
   {journal} {Physical Review A}\ }\textbf {\bibinfo {volume} {101}},\ \bibinfo
  {pages} {053416} (\bibinfo {year} {2020})}\BibitemShut {NoStop}%
\bibitem [{\citenamefont {Shi}\ \emph {et~al.}(2015)\citenamefont {Shi},
  \citenamefont {Robyr}, \citenamefont {Eismann}, \citenamefont {Zawada},
  \citenamefont {Lorini}, \citenamefont {Le~Targat},\ and\ \citenamefont
  {Lodewyck}}]{shiPolarizabilitiesSr872015}%
  \BibitemOpen
  \bibfield  {author} {\bibinfo {author} {\bibfnamefont {C.}~\bibnamefont
  {Shi}}, \bibinfo {author} {\bibfnamefont {J.-L.}\ \bibnamefont {Robyr}},
  \bibinfo {author} {\bibfnamefont {U.}~\bibnamefont {Eismann}}, \bibinfo
  {author} {\bibfnamefont {M.}~\bibnamefont {Zawada}}, \bibinfo {author}
  {\bibfnamefont {L.}~\bibnamefont {Lorini}}, \bibinfo {author} {\bibfnamefont
  {R.}~\bibnamefont {Le~Targat}},\ and\ \bibinfo {author} {\bibfnamefont
  {J.}~\bibnamefont {Lodewyck}},\ }\bibfield  {title} {\bibinfo {title}
  {Polarizabilities of the {{Sr}} 87 clock transition},\ }\href
  {https://doi.org/10.1103/PhysRevA.92.012516} {\bibfield  {journal} {\bibinfo
  {journal} {Physical Review A}\ }\textbf {\bibinfo {volume} {92}},\ \bibinfo
  {pages} {012516} (\bibinfo {year} {2015})}\BibitemShut {NoStop}%
\bibitem [{Note1()}]{Note1}%
  \BibitemOpen
  \bibinfo {note} {ArXiv:2210.16374v1 [physics.atom-ph]}\BibitemShut {NoStop}%
\bibitem [{\citenamefont {Ovsiannikov}\ \emph {et~al.}(2013)\citenamefont
  {Ovsiannikov}, \citenamefont {Pal'chikov}, \citenamefont {Taichenachev},
  \citenamefont {Yudin},\ and\ \citenamefont
  {Katori}}]{ovsiannikovMultipoleNonlinearAnharmonic2013}%
  \BibitemOpen
  \bibfield  {author} {\bibinfo {author} {\bibfnamefont {V.~D.}\ \bibnamefont
  {Ovsiannikov}}, \bibinfo {author} {\bibfnamefont {V.~G.}\ \bibnamefont
  {Pal'chikov}}, \bibinfo {author} {\bibfnamefont {A.~V.}\ \bibnamefont
  {Taichenachev}}, \bibinfo {author} {\bibfnamefont {V.~I.}\ \bibnamefont
  {Yudin}},\ and\ \bibinfo {author} {\bibfnamefont {H.}~\bibnamefont
  {Katori}},\ }\bibfield  {title} {\bibinfo {title} {Multipole, nonlinear, and
  anharmonic uncertainties of clocks of {{Sr}} atoms in an optical lattice},\
  }\href {https://doi.org/10.1103/PhysRevA.88.013405} {\bibfield  {journal}
  {\bibinfo  {journal} {Physical Review A}\ }\textbf {\bibinfo {volume} {88}},\
  \bibinfo {pages} {013405} (\bibinfo {year} {2013})}\BibitemShut {NoStop}%
\bibitem [{\citenamefont {Ovsiannikov}\ \emph {et~al.}(2016)\citenamefont
  {Ovsiannikov}, \citenamefont {Marmo}, \citenamefont {Palchikov},\ and\
  \citenamefont {Katori}}]{ovsiannikovHigherorderEffectsPrecision2016}%
  \BibitemOpen
  \bibfield  {author} {\bibinfo {author} {\bibfnamefont {V.~D.}\ \bibnamefont
  {Ovsiannikov}}, \bibinfo {author} {\bibfnamefont {S.~I.}\ \bibnamefont
  {Marmo}}, \bibinfo {author} {\bibfnamefont {V.~G.}\ \bibnamefont
  {Palchikov}},\ and\ \bibinfo {author} {\bibfnamefont {H.}~\bibnamefont
  {Katori}},\ }\bibfield  {title} {\bibinfo {title} {Higher-order effects on
  the precision of clocks of neutral atoms in optical lattices},\ }\href
  {https://doi.org/10.1103/PhysRevA.93.043420} {\bibfield  {journal} {\bibinfo
  {journal} {Physical Review A}\ }\textbf {\bibinfo {volume} {93}},\ \bibinfo
  {pages} {043420} (\bibinfo {year} {2016})}\BibitemShut {NoStop}%
\bibitem [{\citenamefont {Porsev}\ \emph {et~al.}(2018)\citenamefont {Porsev},
  \citenamefont {Safronova}, \citenamefont {Safronova},\ and\ \citenamefont
  {Kozlov}}]{porsevMultipolarPolarizabilitiesHyperpolarizabilities2018}%
  \BibitemOpen
  \bibfield  {author} {\bibinfo {author} {\bibfnamefont {S.~G.}\ \bibnamefont
  {Porsev}}, \bibinfo {author} {\bibfnamefont {M.~S.}\ \bibnamefont
  {Safronova}}, \bibinfo {author} {\bibfnamefont {U.~I.}\ \bibnamefont
  {Safronova}},\ and\ \bibinfo {author} {\bibfnamefont {M.~G.}\ \bibnamefont
  {Kozlov}},\ }\bibfield  {title} {\bibinfo {title} {Multipolar
  {{Polarizabilities}} and {{Hyperpolarizabilities}} in the {{Sr Optical
  Lattice Clock}}},\ }\href {https://doi.org/10.1103/PhysRevLett.120.063204}
  {\bibfield  {journal} {\bibinfo  {journal} {Physical Review Letters}\
  }\textbf {\bibinfo {volume} {120}},\ \bibinfo {pages} {063204} (\bibinfo
  {year} {2018})}\BibitemShut {NoStop}%
\bibitem [{\citenamefont {Wu}\ \emph {et~al.}(2019)\citenamefont {Wu},
  \citenamefont {Tang}, \citenamefont {Shi},\ and\ \citenamefont
  {Tang}}]{wuDynamicMultipolarPolarizabilities2019}%
  \BibitemOpen
  \bibfield  {author} {\bibinfo {author} {\bibfnamefont {F.-F.}\ \bibnamefont
  {Wu}}, \bibinfo {author} {\bibfnamefont {Y.-B.}\ \bibnamefont {Tang}},
  \bibinfo {author} {\bibfnamefont {T.-Y.}\ \bibnamefont {Shi}},\ and\ \bibinfo
  {author} {\bibfnamefont {L.-Y.}\ \bibnamefont {Tang}},\ }\bibfield  {title}
  {\bibinfo {title} {Dynamic multipolar polarizabilities and
  hyperpolarizabilities of the {{Sr}} lattice clock},\ }\href
  {https://doi.org/10.1103/PhysRevA.100.042514} {\bibfield  {journal} {\bibinfo
   {journal} {Physical Review A}\ }\textbf {\bibinfo {volume} {100}},\ \bibinfo
  {pages} {042514} (\bibinfo {year} {2019})}\BibitemShut {NoStop}%
\bibitem [{\citenamefont {Wu}\ \emph {et~al.}(2023)\citenamefont {Wu},
  \citenamefont {Shi},\ and\ \citenamefont
  {Tang}}]{wuContributionsNegativeenergyStates2023}%
  \BibitemOpen
  \bibfield  {author} {\bibinfo {author} {\bibfnamefont {F.-F.}\ \bibnamefont
  {Wu}}, \bibinfo {author} {\bibfnamefont {T.-Y.}\ \bibnamefont {Shi}},\ and\
  \bibinfo {author} {\bibfnamefont {L.-Y.}\ \bibnamefont {Tang}},\ }\href
  {http://arxiv.org/abs/2301.06740} {\bibinfo {title} {Contributions of
  negative-energy states to the {{E2-M1}} polarizability of the {{Sr}} clock}}
  (\bibinfo {year} {2023}),\ \Eprint {https://arxiv.org/abs/2301.06740}
  {arXiv:2301.06740 [physics]} \BibitemShut {NoStop}%
\bibitem [{\citenamefont {Westergaard}\ \emph {et~al.}(2011)\citenamefont
  {Westergaard}, \citenamefont {Lodewyck}, \citenamefont {Lorini},
  \citenamefont {Lecallier}, \citenamefont {Burt}, \citenamefont {Zawada},
  \citenamefont {Millo},\ and\ \citenamefont
  {Lemonde}}]{westergaardLatticeInducedFrequencyShifts2011}%
  \BibitemOpen
  \bibfield  {author} {\bibinfo {author} {\bibfnamefont {P.~G.}\ \bibnamefont
  {Westergaard}}, \bibinfo {author} {\bibfnamefont {J.}~\bibnamefont
  {Lodewyck}}, \bibinfo {author} {\bibfnamefont {L.}~\bibnamefont {Lorini}},
  \bibinfo {author} {\bibfnamefont {A.}~\bibnamefont {Lecallier}}, \bibinfo
  {author} {\bibfnamefont {E.~A.}\ \bibnamefont {Burt}}, \bibinfo {author}
  {\bibfnamefont {M.}~\bibnamefont {Zawada}}, \bibinfo {author} {\bibfnamefont
  {J.}~\bibnamefont {Millo}},\ and\ \bibinfo {author} {\bibfnamefont
  {P.}~\bibnamefont {Lemonde}},\ }\bibfield  {title} {\bibinfo {title}
  {Lattice-{{Induced Frequency Shifts}} in {{Sr Optical Lattice Clocks}} at the
  10{\textsuperscript{-17}} {{Level}}},\ }\href
  {https://doi.org/10.1103/PhysRevLett.106.210801} {\bibfield  {journal}
  {\bibinfo  {journal} {Physical Review Letters}\ }\textbf {\bibinfo {volume}
  {106}},\ \bibinfo {pages} {210801} (\bibinfo {year} {2011})}\BibitemShut
  {NoStop}%
\bibitem [{\citenamefont {D{\"o}rscher}\ \emph {et~al.}(2022)\citenamefont
  {D{\"o}rscher}, \citenamefont {Klose}, \citenamefont {Palli},\ and\
  \citenamefont {Lisdat}}]{dorscherExperimentalDeterminationE2M12022}%
  \BibitemOpen
  \bibfield  {author} {\bibinfo {author} {\bibfnamefont {S.}~\bibnamefont
  {D{\"o}rscher}}, \bibinfo {author} {\bibfnamefont {J.}~\bibnamefont {Klose}},
  \bibinfo {author} {\bibfnamefont {S.~M.}\ \bibnamefont {Palli}},\ and\
  \bibinfo {author} {\bibfnamefont {C.}~\bibnamefont {Lisdat}},\ }\href
  {http://arxiv.org/abs/2210.14727} {\bibinfo {title} {Experimental
  determination of the {{E2-M1}} polarizability of the strontium clock
  transition}} (\bibinfo {year} {2022}),\ \Eprint
  {https://arxiv.org/abs/2210.14727} {arXiv:2210.14727 [physics]} \BibitemShut
  {NoStop}%
\end{thebibliography}%

\end{document}


\newcommand{\Sr}{${}^{87}$Sr}
\newcommand{\Er}{$E_{r}$}
\renewcommand{\thefigure}{S\arabic{figure}}\makeatother
\renewcommand{\thetable}{S\arabic{table}}\makeatother


\preprint{}
\title{Supplementary Material for Evaluation of lattice light shift at low 10$^{-19}$ uncertainty for a shallow lattice Sr optical clock }
\author{Kyungtae Kim}
\affiliation{JILA, National Institute of Standards and Technology and University of Colorado \\ Department of Physics, University of Colorado, Boulder, Colorado 80309-0440, USA}
\author{Alexander Aeppli}
\affiliation{JILA, National Institute of Standards and Technology and University of Colorado \\ Department of Physics, University of Colorado, Boulder, Colorado 80309-0440, USA}
\author{Tobias Bothwell}
\affiliation{JILA, National Institute of Standards and Technology and University of Colorado \\ Department of Physics, University of Colorado, Boulder, Colorado 80309-0440, USA}
\author{Jun Ye}
\affiliation{JILA, National Institute of Standards and Technology and University of Colorado \\ Department of Physics, University of Colorado, Boulder, Colorado 80309-0440, USA}
\maketitle
\section{Wannier-Stark states and the light shift}
\begin{figure}[h!]
    \includegraphics[width=\columnwidth]{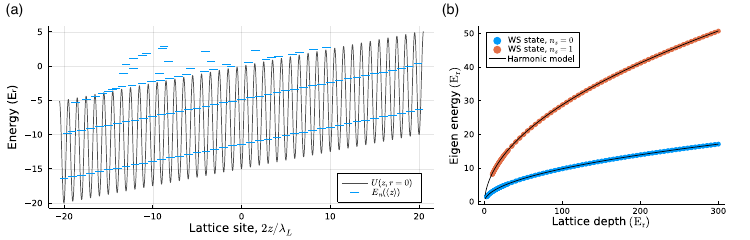}
    \caption{
        (a) Numerically calculated eigenenergy spectrum of the 1D tilted lattice (Eq.~\eqref{eq:potential} with $r=0$) at $U_0= 15E_r$ shown as a function of position expectation value, $\ev{z}$. We use 41 lattice sites with hard edge boundary conditions. The eigenenergies at the edges are distorted due to the boundary. The two lowest eigenstates at the center are shown in Fig. 1(b) in the main text. Higher energy states are untrapped states or artificial bound states due to the boundary condition.
        (b) Comparison of the eigenenergy between Eq.~\eqref{eq:eigenenergy} and the eigenenergy obtained in (a).
    }
    \label{fig:sup_comparison}
\end{figure}

\begin{figure}[h]
    \includegraphics[width=1\columnwidth]{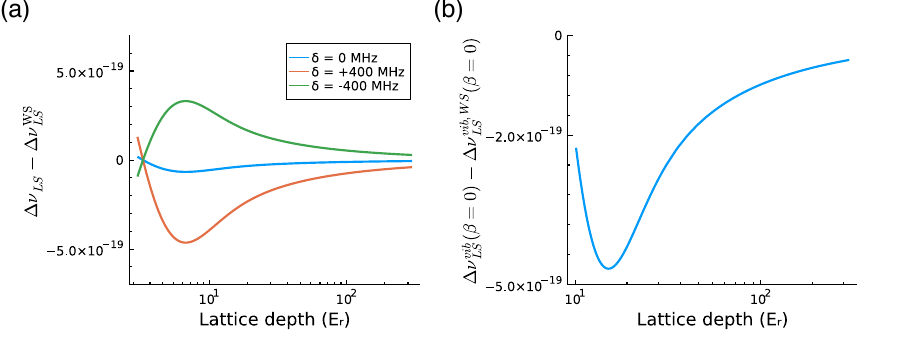}
    \caption{
        The difference between the harmonic model and the WS state model for lattice light shift is calculated. We compare the lattice light shifts based on the analytic form and the one with WS states by subtracting them. This calculation is done without considering the radial temperature ($T_r=0$). The results calculated using WS states are denoted with the superscript ``WS". (a) Detuning and lattice depth dependence. (b) Axial state comparison for the multipolar polarizability ($n_z=1$ and $n_z=0$). Here, $\Delta \nu^{vib}_{LS}(\beta=0) =  \Delta \nu_{LS} ( u, \delta_L=0, n_z=1 )  -  \Delta \nu_{LS} ( u, \delta_L=0, n_z=0 ) $. The calculation is done with $\beta=0$ to only show the effect of $\alpha^{qm}$. 
    }
    \label{fig:Fig2S}
\end{figure}

1D optical lattice potential formed by our cavity, $U$, can be expressed in cylindrical coordinates $z$ and $r$ as
\begin{equation}
    U(z, r) = -U_0\cos^2(kz)e^{-2r^2/w_0^2} + mgz.
    \label{eq:potential}
\end{equation}
Here, $U_0$ is the peak lattice depth, $k$ is the wave vector of the lattice laser, $\omega_0$ is the lattice waist, $m$ is the mass of strontium, and $g$ is the local gravitational acceleration.
The eigenenergy can be calculated from the harmonic basis up to the quartic correction neglecting gravity~\cite{blattRabiSpectroscopyExcitation2009}.
\begin{equation}
    E_{\mathbf{n}}/h \approx \nu_z \left(n_z + \frac{1}{2}\right) + \nu_r (n_x + n_y + 1) - \frac{\nu_{rec}}{2}\left(n_z^2 + n_z + \frac{1}{2}\right) - \nu_{rec}\frac{\nu_r}{\nu_z}(n_x + n_y + 1)\left(n_z + \frac{1}{2}\right),
    \label{eq:eigenenergy}
\end{equation}
 where $\nu_z = 2\nu_{rec}\sqrt{U_0/E_r}$ is the axial trapping frequency, $\nu_r = \sqrt{U_0/m\pi^2w_0^2}$ is the radial trapping frequency, $\nu_{rec} = E_r/h$. $E_{\mathbf{n}}$ is defined with respect to the peak trap depth, $-U_0$. Although this approximation does not take into account any tunneling effects, it captures the eigenenergies of the Wannier-Stark (WS) states as shown in Fig.~\ref{fig:sup_comparison}.
The extended wavefunction of WS states distort the light shift model~\cite{katoriStrategiesReducingLight2015,ushijimaOperationalMagicIntensity2018}. In Fig.~\ref{fig:Fig2S}, we compute differential light shifts without hyperpolarizability with numerically obtained WS states and compare with harmonic model. The effect of extended wavefunctions and higher order corrections causes at most $10^{-19}$ fractional frequency shift. 

For the 1D model $U(z, r=0)$, WS states are not very different from the eigenstates of the harmonic basis with quartic correction in terms of its eigenenergy. Because the perturbation from the tilt ($\sim$1~\si{kHz}) is much smaller than typical band gaps ($\sim$100~\si{kHz}) but large enough to localize the atom (tunneling rate $\sim$10-200~\si{Hz}). However, the extension of the wavefunction has its significance when considering the light-matter coupling, as we observe in reduced Rabi frequency.

\section{Effective radial potential}
Due to the axial-radial coupling of the potential, the effective radial trapping potential is lifted by the motional energy along the axial direction. By collecting the coefficients of $(n_x,\ n_y)$ in Eq.~\eqref{eq:eigenenergy}, we obtain an effective radial trapping frequency, $\bar{\nu}_r^{n_z}$,
\begin{equation}
    \bar{\nu}_r^{n_z}(n_x+n_y) \equiv  \left(\nu_r - \nu_{rec} \frac{\nu_r}{\nu_z}\left(n_z + \frac{1}{2}\right) \right) (n_x+n_y) = 2.44\ \text{Hz} \left(2\sqrt{\frac{U_0}{E_{r}}}-\left(n_z + \frac{1}{2}\right) \right)(n_x+n_y).
    \label{eq:radial_nu}
\end{equation}
Here, we use our cavity waist size, $w_0=260$~\si{\micro\meter}. 

An effective radial potential, $U_{n_z}(r)$ based on Born-Oppenheimer(BO) approximation, derived by Beloy et al.~\cite{beloyModelingMotionalEnergy2020} with Eq.~\eqref{eq:eigenenergy}, is useful with WS states as well. $U_{n_z}(r)$ can be written as
\begin{equation}
    U_{n_z}(r) = -U_0e^{-2r^2/w_0^2} + 2E_{r}\sqrt{\frac{U_0e^{-2r^2/w_0^2}}{E_{r}}} \left(n_z + \frac{1}{2}\right) - \frac{E_{r}}{2}\left(n_z^2 + n_z + \frac{1}{2}\right).
    \label{eq:Unz}
\end{equation}
The second and third term corrects the radial trapping potential by the axial motion's energy and the correction depends on the axial quantum number, $n_z$. We expand $U_{n_z}(r)$ in terms of small ($r^2/w_0^2$) to extract the effective radial trapping frequency.
\begin{equation}
    \begin{aligned}
        U_{n_z}(r) &\approx -U_0+h\nu_z\left(n_z + \frac{1}{2}\right) + \left( \frac{2U_0}{w_0^2} - \frac{h\nu_z}{w_0^2} \left(n_z + \frac{1}{2}\right) \right)r^2 \\
        & =  -U_0+h\nu_z\left(n_z + \frac{1}{2}\right)  + \frac{1}{2}m\frac{4U_0}{mw_0^2} \left( 1 - \frac{1}{2}\sqrt{\frac{E_{r}}{U_0}}\left(n_z + \frac{1}{2}\right) \right)^2r^2.
    \end{aligned}
\end{equation}
This gives us a consistent result with Eq.~\eqref{eq:radial_nu}.

\begin{figure}[h!]
    \includegraphics[width=0.8\columnwidth]{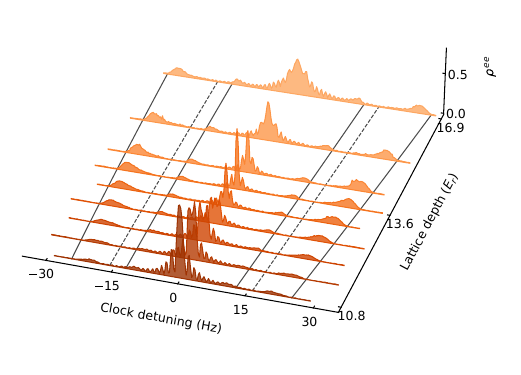}
    \caption{
        Resolved radial sidebands for $\ket{n_z=1}$. A profound change in the central part of the spectrum reflects the reduction of he Rabi frequency (see Fig. 2 in the main text). The laser intensity is set to be $\pi$-pulse for 13~\Er{} and kept the same for other depths. The solid (dashed) guidelines are expected maximum radial trapping frequency and its harmonics with (without) considering axial-radial coupling, based on Eq.~\eqref{eq:radial_nu}
    }
    \label{fig:radial_sideband_nz1}
\end{figure}
In Fig.~2 in the main text, we observe the enhanced tunneling rate for $\ket{n_z=1}$ results in more than tenfold reduction of the carrier Rabi frequency as $U_0$ is reduced to near $13$\Er{}, at which point we increase the clock laser's power by 100 to compensate for the reduced coupling. The change of the carrier and WS+$i$ Rabi frequency is apparent at the central part of the spectra in Fig.~\ref{fig:radial_sideband_nz1}. 

The resolved radial sidebands also evolve dramatically.  The sidebands shift towards larger frequency as we increase $U_0$. As higher order corrections broaden the radial frequencies, Eq.~\eqref{eq:radial_nu} captures the highest frequency of the radial sidebands, indicated by the solid gray guidelines in Fig.~\ref{fig:radial_sideband_nz1}. The gray dashed guideline shows $\sqrt{U_0}$ scaling. The difference between the two lines highlights the influence of axial-radial coupling and clarifies the decrease in $T_r$ shown in Fig.~1(d) of the main text, when a fixed spatial profile is maintained during excitation. In order to preserve the spatial distribution after the excitation to $\ket{n_z=1}$, where the atoms experience decreased confinement, the temperature of the atoms must be lowered.

Interestingly, the second radial sidebands can have a larger amplitude than the first sideband for $\ket{n_z=1}$. Near 13\Er{}, where Rabi frequency becomes exponentially sensitive to the lattice intensity, the Gaussian profile of the lattice laser introduces radially varying Rabi frequency. Therefore, the Gaussian mode of the lattice laser inherently generates coupling to the radial modes, even with a plane wave clock laser.

\begin{figure}[h!]
    \includegraphics[width=0.7\columnwidth]{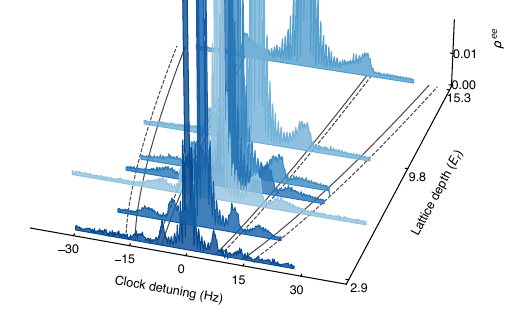}
    \caption{
        Zoomed-in view of the radial sideband of the axial ground state. The laser intensity is set to be $\pi$-pulse for 9.8~\Er{} and kept the same for other depths. The solid (dashed) guidelines are expected maximum radial trapping frequency and its harmonics with (without) considering axial-radial coupling, based on Eq.~\eqref{eq:radial_nu}
    }
    \label{fig:radial_sideband_nz0}
\end{figure}

The coupling strength to the radial sideband is suppressed for the state $\ket{n_z=0}$, compared to $\ket{n_z=1}$. Figure~\ref{fig:radial_sideband_nz0} displays the radial sideband spectrum for the $\ket{n_z=0}$ state, where the dominant sidebands are the first order ones. As the lattice depth is reduced, the relative coupling strength to the second order sideband increases, due to the enhanced sensitivity of the Rabi frequency to the lattice depth.

\section{Rabi frequency extraction simulation}
\begin{figure}
    \includegraphics[width=0.7\columnwidth]{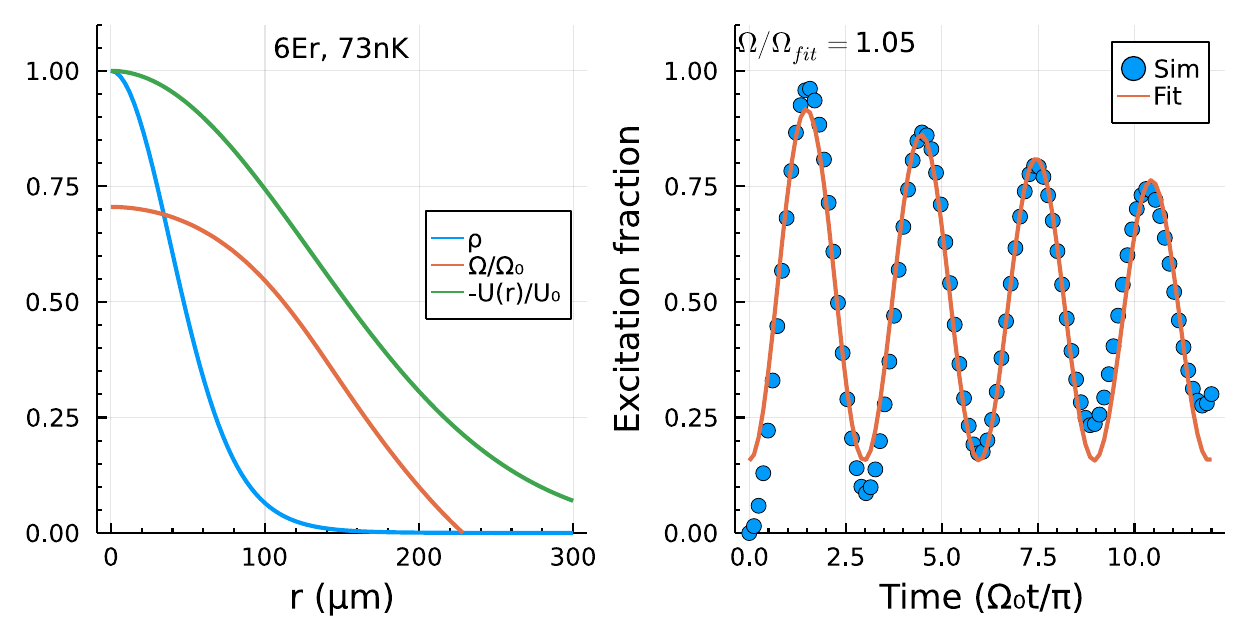}
    \caption{
        Rabi frequency extraction simulation at 6~\Er{} for $\ket{n_z=0}$. $\rho$ is the normalized radial density of the atoms, $\Omega$ is the Rabi frequency as a function of $r$, $\Omega_0$ is the free space Rabi frequency, and $U(r)$ is the potential.
    }
    \label{fig:sup_rabi}
\end{figure}

Based on Eq.~\eqref{eq:Unz}, we simulate the effect of inhomogeneous Rabi frequency on a global Rabi flopping measurement. We spatially average the excitation fraction with Boltzmann weighted Rabi flopping to get global excitation fraction at a given time, 
\begin{equation}
    \rho^{ee}(t) = \left. \int_0^{\infty}{\exp(-\frac{U_{n_z}(r)}{k_B T_r}) \sin(\frac{\Omega(r) t}{2})rdr} \right/ \int_0^{\infty}{\exp(-\frac{U_{n_z}(r)}{k_B T_r})rdr}.
\end{equation}
We fit measured $\rho^{ee}(t)$ to a function $\rho^{ee} = A\sin(\Omega_{fit}t/2)\exp(-Bt)+C$. All capital letters are the fitting coefficient, and we use $\Omega_{fit}$ to estimate the actual peak Rabi frequency ($\Omega(r=0)$). Based on least squares fitting to the Doppler profile, the radial temperature $T_r$ follows
\begin{equation}
    T_r(u_0) = A\left(u_0 -B\right)^{\kappa}~\si{nK},
\end{equation}
with $A = 31.6(1.4)$, $B = 2.2(0.3)$, $\kappa = 0.58(0.01)$. 
In most cases, the extracted Rabi frequencies are matched to $\Omega(r=0)$ within 10~\%. The Rabi flopping is more complicated for the cases where the inhomogeneity of the Rabi frequency is prominent ($ 12-17 E_{r} $ for $n_z =1$ state). The simulation at 6$E_r$ is shown in Fig.~\ref{fig:sup_rabi}.

\section{Extended data and systematic error at the very shallow lattice depth}
\begin{figure}[h!]
    \includegraphics[width=0.7\columnwidth]{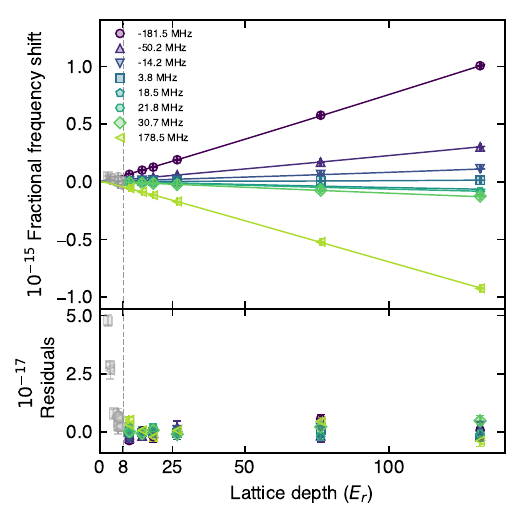}
    \caption{
        Extended data for Fig. 3 in the main text. The solid lines represent the fit without data points below 8\Er{}. The gray dashed line marks the cutoff. The model is unable to accurately describe the data at very shallow lattice depths.
    }
    \label{fig:Fig3_extended}
\end{figure}

\begin{figure}[h!]
    \includegraphics[width=0.5\columnwidth]{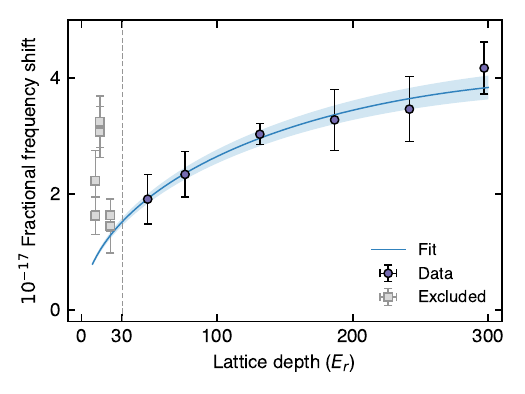}
    \caption{
        Extended data for Fig. 4 in the main text. The solid lines are the fit without data points below 30\Er{}. The gray dashed line indicates the depth cutoff. As in the case of Fig.~\ref{fig:Fig3_extended}, the model does not capture the shallow depth data.
    }
    \label{fig:Fig4_extended}
\end{figure}

Figure \ref{fig:Fig3_extended} and \ref{fig:Fig4_extended} shows extended data including the data points we exclude from the fit. We also present the contribution from the systematic and statistical uncertainties to the final uncertainties in Table~\ref{tab:1S}.

\begin{table}
    \centering
    \def\arraystretch{1.3}%
    \begin{tabular}{ c c c }
        \hline\hline
        Quantity & Value\\
        \hline
        $\partial_{\nu} \tilde{\alpha}^{E1}/ h$             &  $1.859(0.004)_{\text{stat}}(0.003)_{\text{sys}} \times 10^{-11}$        \\
        $\nu^{E1}$ (\si{\mega\hertz})                         & $368,554,825.9(0.4)_{\text{stat}}(0.002)_{\text{sys}}$                           \\
        $\tilde{\alpha}^{qm}/h$ (\si{\milli\hertz})         & $-1.24(0.05)_{\text{stat}}(0.001)_{\text{sys}}$                                               \\
        $\tilde{\beta}/h$ (\si{\micro\hertz})                            & $-0.510(0.037)_{\text{stat}}(0.001)_{\text{sys}}$                                              \\
        \hline\hline
    \end{tabular}
    \caption{
        Extended Tab.~1 in the main text. Statistical error reflects propagated uncertainties of control parameters through orthogonal distance regressions, while systematic uncertainties originate from radial temperature variations.
        }
    \label{tab:1S}
\end{table}

The effect of the extended wavefunction and the corresponding loss (coupling to upper, untrapped bands) at low lattice depth could introduce systematic effects. We observe significant disagreement between the data and the model at particularly low lattice depths ($\leq 6 $\Er{} for $(u, \delta_L)$ modulation, $\leq 30$\Er{} for $n_z$ modulation). The reduced chi-squared of the model fit rapidly increases as we include the low lattice depth data points. The deviation from the model depends only on the lattice depth, insensitive to the change of $\delta_L$ by 180~\si{MHz}, ruling out the lattice light shift. 
Modifying the clock duration, initial state (preparing in $^3P_0$), and the bias field similarly have no effect.

\begin{figure}[h]
    \includegraphics[width=0.99\columnwidth]{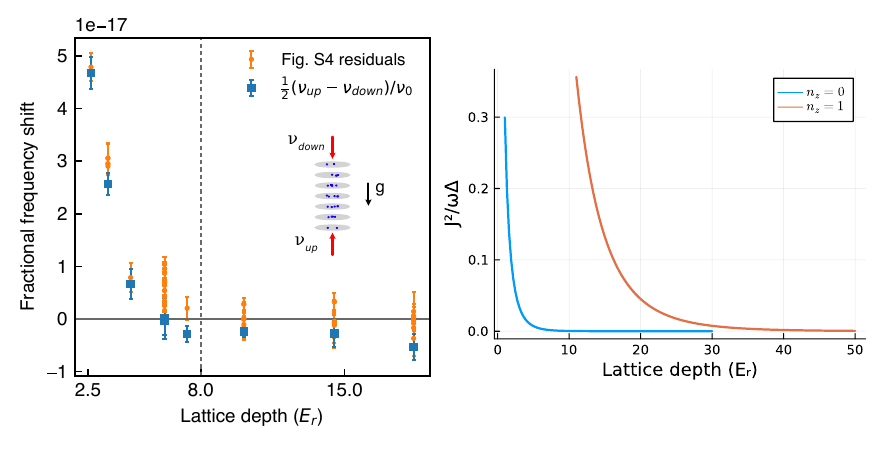}
    \caption{
        Left: frequency comparison of upward and downward propagating clock interrogation. We use upward clock interrogation for all the lattice light shift measurements. The frequency shift matches the trend and the magnitude to the fit residual of Fig.~\ref{fig:Fig4_extended}. The clock direction-related frequency shift sharply rises at the very shallow lattice depth, not captured by the lattice light shift model. The gray dashed line indicates the depth cutoff based on the reduced chi-squared of the fit of Fig.~\ref{fig:Fig4_extended}. The offset from zero (blues) is presumably from technical imperfections of the fiber noise cancelation system of the downward clock laser.
        Right: Illustration of the dramatic increase of the effective tunneling rate across the bandgap as the lattice strength decreases, for both the axial ground state and the first excited state. The tunneling rate of either the ground or the first excited band is represented by the parameter $J$, and $\Delta$ represents the energy difference between neighboring lattice sites due to the gravity. $J$ and $\omega$ are calculated for a case where there is no tilt.
    }
    \label{fig:top_down_comp}
\end{figure}

However, comparing the clock frequency shift between upward and downward clock laser propagation under the same condition [Fig.~\ref{fig:top_down_comp}, left] results in frequency differences at twice the deviation from the model fit. In other words, the deviation from the model changes sign as we reverse the propagation direction of the clock laser (with respect to gravity), but the magnitude remains the same. Although the measurement use different transitions ($m_F = \pm 9/2 \rightarrow \pm 7/2$) and a less ideal downward clock laser, this measurement provides a strong evidence for a systematic frequency shift associated with non-stationary states at the low lattice depth. We do not have a quantitative model for the shift, but we present a suggestive plot of the effective tunneling rate in the right panel of Fig.~\ref{fig:top_down_comp}. The Doppler shift of the atoms slowly moving upwards (15~\si{nm/s} for $5\times 10^{-17}$) also can explain the deviation, but we do not find any source of such movement. Further theoretical investigations will follow in future work. Based on the observations and the behavior of $\chi^2$ of the fit, we decide to exclude data points with $\leq 6 $\Er{} for $(u, \delta_L)$ modulation, $\leq 30$\Er{} for $n_z$ modulation from the model fit.

\begin{figure}[h!]
    \includegraphics[width=0.85\columnwidth]{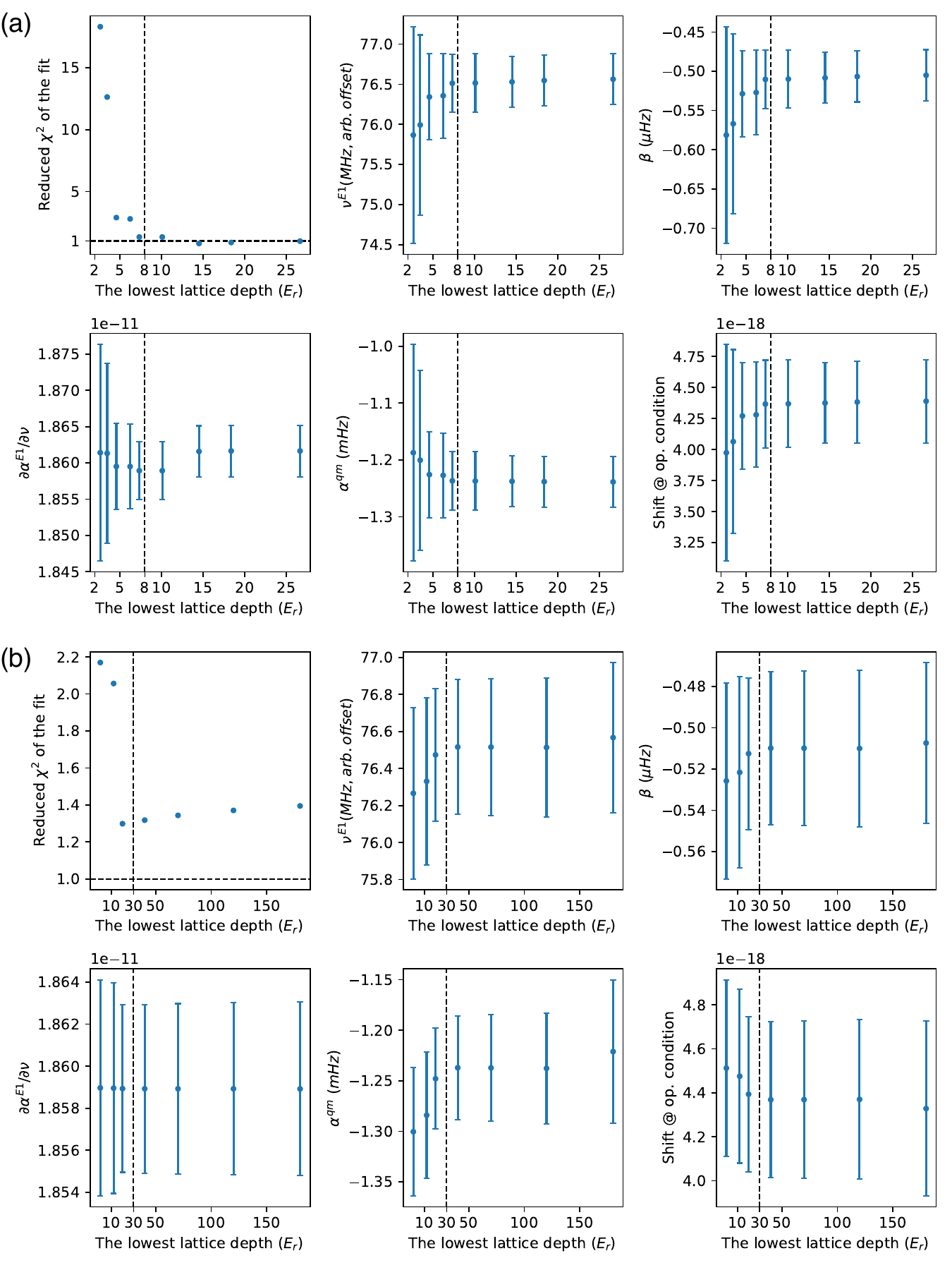}
    \caption{
        Lattice depth cutoff test. The figure presents a test of the depth cutoff for (a) $(u, \delta_L)$ modulation cutoff test with fixed 30\Er{} cutoff for $n_z$ modulation, and (b) $n_z$ modulation cutoff test with fixed 8\Er{} cutoff for $(u, \delta_L)$ modulation. The term ``Shift @ op. condition" refers to the light shift calculated under an operational condition, specifically $u = 10(0.2)E_r, \delta_L=-14.7(0.1)~\si{MHz}, n_z=0(0.03)$. The cutoff values are indicated by the dotted lines.
    }
    \label{fig:depth_versus_chisq}
\end{figure}

We test the stability of the fitting coefficients with respect to the depth cutoff. Figure~\ref{fig:depth_versus_chisq} displays the results of the fitting parameters for various depth cutoffs. Most of the parameters exhibit rapid increase in uncertainty for shallow lattice depths, although the values remain consistent with our quoted results.

\section{Vector polarizability contribution}
We examine the frequency splitting between opposite spin states as a function of lattice depth. By fitting our data to the model described in \cite{shiPolarizabilitiesSr872015}, we determine a lattice light ellipticity of 0.136(2) and an angle of 89.96(3) degrees between the lattice light k-vector and the bias field.
This analysis also reveals an additional light shift term, proportional to $u^2$, that arises from the ellipticity and the vector-tensor coupling. Despite its presence, the fractional shift remains below $10^{-19}$ for all lattice depths and bias fields used in our measurement, and even falls below $10^{-20}$ under our operational conditions.

\section{Two data analysis methods}
In this section, we outline the difference between the original data analysis method~\footnote{arXiv:2210.16374v1 [physics.atom-ph]} and the improved version that is more directly connected to the light shift model and produces more robust results over the entire relevant range of the lattice depth.

 In the experiment, two conditions (A and B) were interleaved during data collection to reject common systematic errors such as clock laser frequency drift. The original method disregards information from one of the control parameters (condition B), and instead uses a fitting coefficient ($C$) to represent the shift of B. In this case we can write the light shift for $u, \delta_L$ modulation to be
 \begin{equation}
    \begin{gathered}    
        \Delta \nu_{LS}( u^A, \delta_L^A , 0)  -  \Delta \nu_{LS} ( u^B, \delta_L^B , 0) =  \Delta \nu_{LS}( u^A, \delta_L^A , 0) + C \\
        =  \frac{1}{2}\left( \frac{\partial \tilde{\alpha}^{E1}}{\partial \nu} \delta_L^A - \tilde{\alpha}^{qm} \right) (u^A)^{1/2}  - \left[ \frac{\partial \tilde{\alpha}^{E1}}{\partial \nu} \delta_L^A + \frac{3}{4}\tilde{\beta}\right] u^A + \tilde{\beta} (u^A)^{3/2} - \tilde{\beta} (u^A)^2 + C.
        \label{eqS:light_shift}
    \end{gathered}
\end{equation}
For the case of axial state ($n_z$) modulation, we use the following form.
\begin{equation}
    \Delta \nu_{LS}^{\text{vib}}(u, \delta_L) = \Delta \nu_{LS}(u, \delta_L, 1)  - \Delta \nu_{LS}(u, \delta_L, 0)
    \label{eqS:light_shift_vib}
   \end{equation}
We can fit the data to this model and estimate coefficients including $C$. We see that $C =  -  \Delta \nu_{LS} ( u^B, \delta_L^B , 0)$ represents the total light shift at the condition B. Therefore, by fitting $C$, we evaluate the total lattice light shift at this condition. 

The new data analysis method incorporates information from both conditions A and B to determine the fitting coefficients. This eliminates the requirement for the coefficient $C$, which was previously used as a reference point. The inclusion of information from both A and B leads to a reduction in the number of sensitive fit parameters, improving accuracy and precision of the results. Although the total number of coefficients remains the same with the inclusion of $\tilde{\beta}$, it has a small contribution at the shallow lattice depth, effectively reducing the number of fit coefficients used in the final evaluation.

The old analysis method uses two separate fitting routines (see also \cite{ushijimaOperationalMagicIntensity2018}). The first step is to extract $\tilde{\alpha}^{qm}$ from equation~\eqref{eqS:light_shift_vib} with $\tilde{\beta}$ fixed to the average of previously known values. Using this fitted $\tilde{\alpha}^{qm}$, we fit equation ~\eqref{eqS:light_shift} to extract $\partial_{\nu}\alpha^{E1}$, $\nu^{E1}$, $C$. With the new method, the data analysis is greatly simplified. A comprehensive global fitting routine is employed, which simultaneously fits all coefficients, including both $\tilde{\alpha}^{qm}$ and $\tilde{\beta}$. This leads to a more comprehensive and self-contained analysis. Furthermore, the new method provides a statistically consistent result when estimating the light shift of condition B. The result of the old analysis method is summarized in Tab.~\ref{tab:old_result}

\begin{table}
    \centering
    \def\arraystretch{1.3}%
    \begin{tabular}{ c c c }
        \hline\hline
        Quantity & Fig.~3 & Fig.~3, 4\\
        \hline
        $\partial_{\nu} \tilde{\alpha}^{E1}/ h$ &  $1.855(5)\times 10^{-11}$ & $1.855(5)\times 10^{-11}$ \\
        $\nu^{E1}$ (\si{\mega\hertz}) & $368,554,825.8(1.5)$ & $368,554,825.2(0.2)$ \\
        $\tilde{\alpha}^{qm}/h$ (\si{\milli\hertz}) & $-0.9(0.8)$ & $-1.25(5)$ \\
        $C\ (10^{-18})$  & $-9.2(2.7)$& $-10.3(0.5)$\\
        \hline\hline
    \end{tabular}
    \caption{
        Summary of the light shift characterization with the old analysis method. The second column present the fit result only with ($\delta_L$, $u$) modulation data set, and the third column shows the result with using $\tilde{\alpha}^{qm}$ measured with $n_z$ modulation.
        }
    \label{tab:old_result}
\end{table}

\section{Summary of reported multipolar polarizability}
\begin{figure}[h!]
    \input{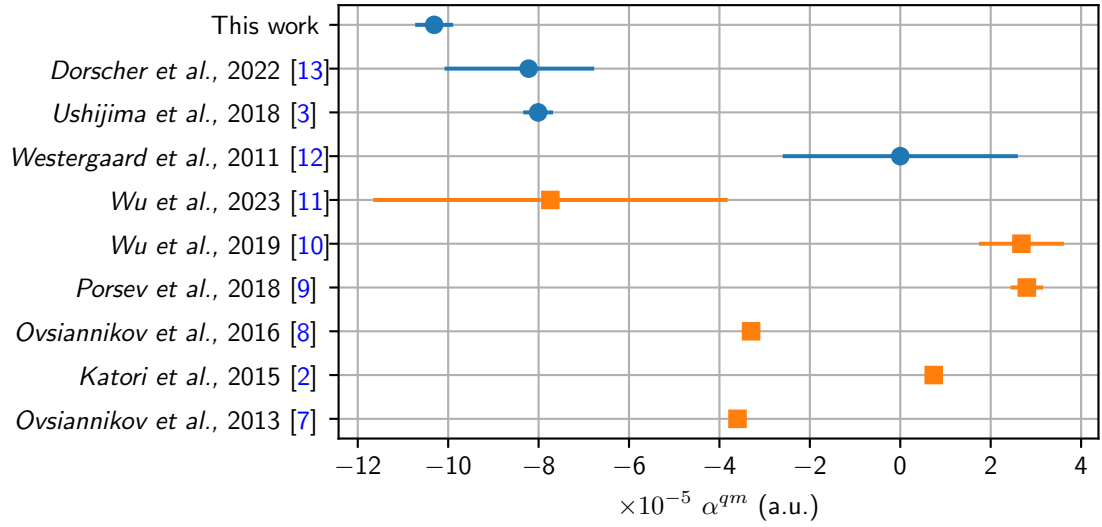}
    \caption{
        The reported values of $\alpha^{qm}$ in atomic units. The blue circles represent experimental results and the orange squares represent theoretical calculations.
    }
    \label{fig:alpha_qm_summary}
\end{figure}

Figure~\ref{fig:alpha_qm_summary} shows the reported values of $\alpha^{qm}$ in atomic units. For this work, $\alpha^{E1} = 288.6$ a.u. is used for the unit conversion.

\clearpage
\bibliography{main}